\documentclass[]{aa}  

\usepackage{graphicx}
\usepackage{fixltx2e}
\usepackage{txfonts}
\usepackage{natbib}
\bibpunct{(}{)}{;}{a}{}{,} 

\begin{document}

   \title{\emph{Herschel} spectroscopic observations of the compact obscured nucleus in Zw 049.057\thanks{\emph{Herschel} is an ESA space observatory with science instruments provided by European-led Principal Investigator consortia and with important participation from NASA.}}

   \subtitle{}

   \author{N. Falstad \inst{1}
          \and
          E. Gonz\'alez-Alfonso \inst{2}
          \and
          S. Aalto \inst{1}
          \and
          P.P. van der Werf \inst{3}
          \and
          J.~Fischer \inst{4}
          \and
          S.~Veilleux\inst{5}
          \and
          M.~Mel{\'e}ndez\inst{5}
          \and
          D. Farrah\inst{6}
          \and
          H.~A.~Smith\inst{7}
\fnmsep
          }

   \institute{Department of Earth and Space Sciences, Chalmers University of Technology, Onsala Space Observatory,
              439 92 Onsala, Sweden \\
              \email{niklas.falstad@chalmers.se}
         \and
             Universidad de Alcal\'a de Henares, 
             Departamento de F\'{\i}sica, Campus Universitario, E-28871 Alcal\'a de
             Henares, Madrid, Spain  
         \and
             Leiden Observatory, Leiden University,
             P.O.\ Box 9513, NL-2300 RA Leiden, The Netherlands
         \and
             Naval Research Laboratory, Remote Sensing Division, 4555
             Overlook Ave SW, Washington, DC 20375, USA
         \and
             Department of Astronomy, University of Maryland, College Park, MD 20742, USA
         \and
             Department of Physics, Virginia Tech, Blacksburg, VA 24061, USA
         \and
             Harvard-Smithsonian Center for Astrophysics, 60 Garden Street, Cambridge, MA 02138, USA
             }

  \date{}
 
  \abstract
   {The luminous infrared galaxy \object{Zw~049.057} contains a compact obscured nucleus where a considerable amount of the galaxy's luminosity is generated. This nucleus contains a dusty environment that is rich in molecular gas. One approach to probing this kind of environment and to revealing what is hidden behind the dust is to study the rotational lines of molecules that couple well with the infrared radiation emitted by the dust.}
   {We probe the physical conditions in the core of Zw~049.057 and establish the nature of its nuclear power source (starburst or active galactic nucleus).}
   {We observed Zw~049.057 with the Photodetector Array Camera and Spectrometer (PACS) and the Spectral and Photometric Imaging Receiver (SPIRE) onboard the \emph{Herschel} Space Observatory in rotational lines of H$_{2}$O, H$_{2}^{18}$O, OH, $^{18}$OH, and [O~I]. We modeled the unresolved core of the galaxy using a spherically symmetric radiative transfer code. To account for the different excitation requirements of the various molecular transitions, we use multiple components and different physical conditions.}
   {We present the full high-resolution SPIRE FTS spectrum of Zw~049.057, along with relevant spectral scans in the PACS range. We find that a minimum of two different components (nuclear and extended) are required in order to account for the rich molecular line spectrum of Zw~049.057. The nuclear component has a radius of $10-30$~pc, a very high infrared surface brightness ($\sim10^{14}$~L$_{\sun}~$kpc$^{-2}$), warm dust ($T_{\mathrm{d}}>100$~K), and a very large H$_{2}$ column density (N$_{\mathrm{H_{2}}}=10^{24}-10^{25}$~cm$^{-2}$). The modeling also indicates high nuclear H$_{2}$O ($\sim 5 \times 10^{-6}$) and OH ($\sim 4 \times 10^{-6}$) abundances relative to H$_{2}$ as well as a low $^{16}$O/$^{18}$O-ratio of $50-100$. We also find a prominent infall signature in the [O~I] line. We tentatively detect a $500$~km~s$^{-1}$ outflow in the H$_{2}$O $3_{13}\!\rightarrow\!2_{02}$ line.}
   {The high surface brightness of the core indicates the presence of either a buried active galactic nucleus or a very dense nuclear starburst. The estimated column density towards the core of Zw~049.057 indicates that it is Compton-thick, making a buried X-ray source difficult to detect even in hard X-rays. We discuss the elevated H$_{2}$O abundance in the nucleus in the context of warm grain and gas-phase chemistry. The H$_{2}$O abundance is comparable to that of other compact (ultra-)luminous infrared galaxies such as NGC~4418 and Arp~220 - and also to hot cores in the Milky Way. The enhancement of $^{18}$O is a possible indicator that the nucleus of Zw~049.057 is in a similar evolutionary stage as the nuclei of Arp~220 - and more advanced than NGC~4418. We discuss the origin of the extreme nuclear gas concentration and note that the infalling gas detected in [O I] implies that the gas reservoir in the central region of Zw~049.057 is being replenished. If confirmed, the H$_{2}$O outflow suggests that the nucleus is in a stage of rapid evolution.}
 

   \maketitle
%

\section{Introduction}
Many luminous infrared galaxies (LIRGs) host compact obscured nuclei (CONs) where a bolometric luminosity L$_{\rm bol}>10^9$ L$_{\sun}$\ emerges from inside a core of diameter $d<100$~pc that is obscured by dust corresponding to a high visual extinction $A_{\rm v}>1000$~mag. The nature of the nuclear power source is thus hidden from examination with conventional methods like optical and infrared (IR) lines and the Compton-thick shroud may render even X-rays unusable. It is important to determine if it is an accreting black hole or a compact starburst that powers the nuclear activity since this greatly affects our understanding of galaxy evolution. It has been suggested that a large portion of highly obscured Compton-thick active galactic nuclei (AGN) are missed by X-ray surveys, a problem that may be even worse for low-luminosity AGN \citep[e.g.,][]{lus13}. 

Furthermore, more than 50\% of the star formation at high redshifts may be obscured \citep{cha05,war11}. Obscured star formation can be linked to the assembly of stellar mass, with deeper potential wells in massive galaxies providing dense, heavily obscured environments resulting in rapid star formation \citep[e.g.,][]{iba13}.

The dust in obscured galaxies may have a high optical depth for a wide span of wavelengths. To study their nuclear structure, dynamics, and physical conditions, we thus need a tracer that can probe deep into the dust shroud. Molecules like water (H$_{2}$O) and hydroxyl (OH) couple very well with the IR-field and can reach high abundances in warm embedded regions, making them ideal probes of physical conditions in dust-enshrouded galaxies as shown in previous studies of \object{Mrk 231}, \object{Arp 220}, and \object{NGC 4418}, among others \citep{fis10,fis14,gon04,gon08,gon10,gon12,gon14,gon14a}. The OH molecule has also been proven to trace massive molecular outflows in ultraluminous infrared galaxies (ULIRGs) \citep{fis10,stu11,vei13,spo13,gon14a}. Recent studies have found H$_{2}$O in high-redshift sources \citep{imp08,omo11,van11,bra11}. Understanding the role of H$_{2}$O in the local Universe will thus aid any interpretation of such high-redshift observations.

In the near Universe, a small sample of CONs have been identified with deep mid-IR silicate absorption and hot optically thick dust cores \citep[e.g.,][]{aal12,cos10,cos13,gon15,sak10,sak13}. We have used the Photodetector Array Camera and Spectrometer \citep[PACS;][]{pog10} and the Spectral and Photometric Imaging Receiver \citep[SPIRE;][]{gri10} onboard the \emph{Herschel} Space Observatory \citep{pil10} to observe H$_{2}$O and OH in the CON Zw 049.057, which is an OH megamaser \citep{baa87} and has moderate IR luminosity \citep[$L_{\mathrm{IR}}\approx 1.8 \times 10^{11}$~L$_{\odot}$][]{san03}, a very rich molecular spectrum in the far-IR and submillimeter, and a compact molecular distribution \citep{pla91}. Its NICMOS image \citep{sco00} reveals a narrow dust feature emerging along the minor axis from the obscured nucleus. \citet{sco00} also suggest that the feature might be due to a nuclear absorbing cloud that blocks the light along the minor axis. Based on radio observations, \citet{baa06} classified Zw~049.057 as an AGN, but it is only weakly detected in X-rays \citep{leh10}. Mid-infrared observations with \emph{Spitzer} have revealed strong silicate absorption \citep{per10} and far-IR \emph{ISO} observations indicated that the LIRGs Zw~049.057, NGC~4418, and IC~860 are all [C~II] deficient \citep{mal97}.

Our interest in this galaxy was originally sparked by the SPIRE FTS observations (see Sect. \ref{sec:full_spire}), which revealed a rich molecular spectrum with H$_{2}$O rotational lines that are comparable in strength to and, in some cases, stronger than the CO rotational lines. Modeling then raised the possibility that Zw~049.057 has an extremely obscured nucleus with an embedded power source, possibly a young AGN. Very little is known about the obscured growth of AGNs, so that finding and identifying such objects is an important step in our understanding of galaxy evolution.

The \emph{Herschel} observations are described in Sect. \ref{sec:observations}, and our models are shown in Sect. \ref{sec:models}. The model results are discussed in Sect. \ref{sec:discussion}, and our main conclusions are summarized in Sect. \ref{sec:conclusions}.


\section{Observations and results}
\label{sec:observations}
Zw~049.057 was observed using the PACS and SPIRE spectrometers onboard the \emph{Herschel} Space Observatory. The PACS observations\footnote{The observation identification numbers (OBSIDs) for the PACS observations are 1342248365-1342248368} were conducted as part of the Hermolirg OT2 project (PI: Gonz\'alez-Alfonso) on 2012 July 20 in high spectral sampling range spectroscopy mode using first and second orders of the grating. The data reduction was done with the \emph{Herschel} Interactive Processing Environment \citep[HIPE;][]{ott10} version 12.1.0 using the PACS background normalization pipeline for short range scans. At a distance of $56$~Mpc, the far-IR nuclear emission of Zw~049.057 is spatially unresolved in the central $9.1''$ ($\sim2.5$~kpc) spaxel of the $5$~x~$5$ spaxel integral field array. Since the point spread function (PSF) is larger than the central spaxel, to obtain the full flux we extracted the central spectrum using the point source correction task in HIPE 12.1.0. In each spectral range scan the continuum was then subtracted using polynomials of third order or lower and the lines were fitted with Gaussian profiles. Line fluxes, measured and inferred intrinsic Gaussian line widths, and continuum levels for the PACS observations are listed in Table \ref{tab:pacslines}.

The corresponding SPIRE observations\footnote{The OBSID of the SPIRE observations is 1342203077} were conducted as part of the OT key program Hercules (PI: van der Werf) on 2011 January 6 in high spectral resolution, single pointing, and sparse image sampling mode. A total of 108 repetitions (216 FTS scans) were carried out, giving an on-source integration time of $14386$ s. The data were processed using the standard point-source pipeline in HIPE 12.1.0. After subtracting the continuum with a third order polynomial in each of the two bands, we measured the fluxes of the identified ($\geq 3\sigma$ level) lines by simultaneously fitting the lines with line profiles consisting of a convolution of the FTS full resolution instrumental response (a sinc function) with the best fit Gaussian line profile of the emission from the galaxy. The full SPIRE FTS spectrum is presented in Sect. \ref{sec:full_spire}. Line fluxes and continuum levels for the H$_{2}$O lines observed with SPIRE are listed in Table \ref{tab:spirelines}. In addition to the SPIRE observations, the Hercules program also included PACS observations of the [O I] $63$~$\mu$m line\footnote{The OBSID for the PACS observations of [O I] $63$~$\mu$m is 1342190374}. These observations were conducted on 2010 February 11. 

\begin{figure}
    \centering
    \includegraphics[width=8.0cm]{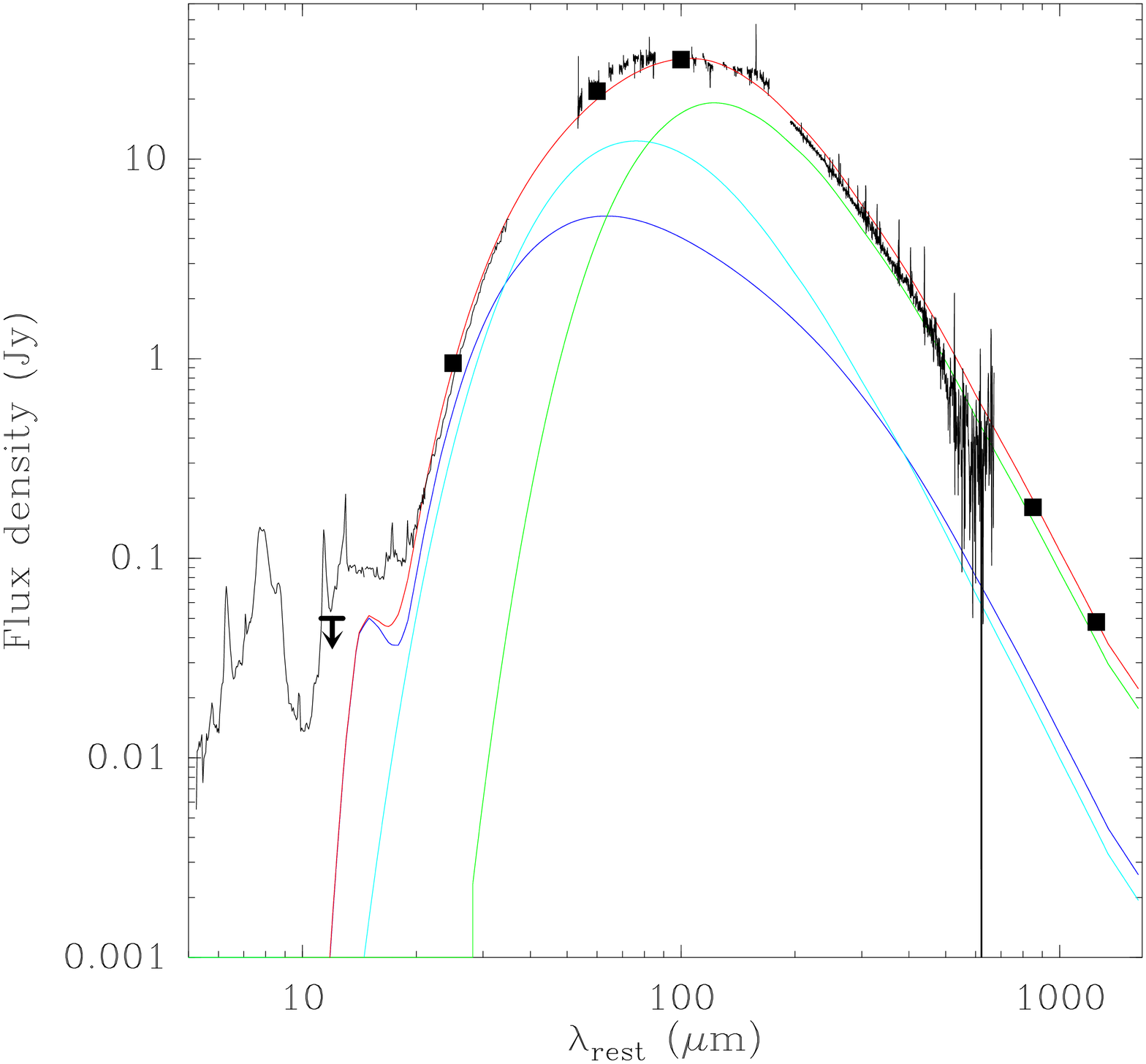}
    \caption{Spectral energy distribution of Zw~049.057 from mid-IR to millimeter wavelengths. Data from \emph{Herschel}/PACS, \emph{Herschel}/SPIRE, and \emph{Spitzer}/IRS are shown. The IRAS data points at $25$, $60$, and $100$~$\mu$m as well as the upper limit at $12$~$\mu$m come from \citet{san03}, the (sub)millimeter points at $850$ and $1250$~$\mu$m are from \citet{dun00} and \citet{car92}, respectively. The models discussed in Sect. \ref{sec:models} are included where the blue and light blue curves represent the core and outer components, respectively. The green curve is an extra component to fit the continuum. Red is the total model.}  
    \label{fig:continuum}
   \end{figure}

The spectral energy distribution (SED) of the source is shown in Fig. \ref{fig:continuum}, including the \emph{Spitzer}/IRS spectrum \citep{arm09}, and photometric data points from IRAS \citep{san03} and SCUBA \citep{dun00}. Note the strong silicate absorption at $9.7$ and $18$~$\mu$m reported by \citet{per10}.

\begin{figure}[ht]
    \centering
    \includegraphics[width=8.0cm]{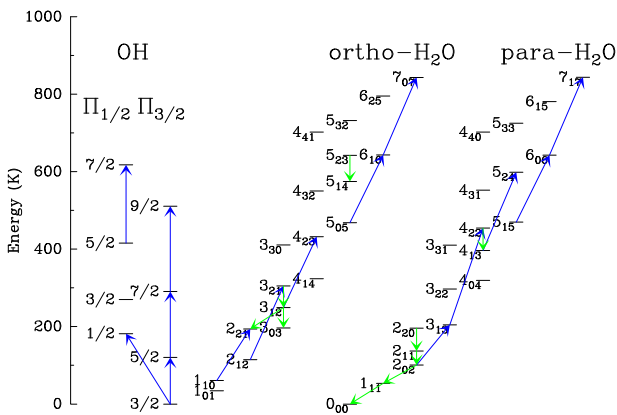}
    \caption{Energy level diagrams of OH and H$_{2}$O (ortho and para). Blue arrows indicate lines observed with PACS and green arrows denote lines observed with SPIRE.}  
    \label{fig:energy_diagram}
   \end{figure}

Spectroscopic parameters for OH and H$_{2}$O (both ortho and para) used for line identification and radiative transfer modeling were taken from the JPL \citep{pic98} and CDMS \citep{mul01,mul05} catalogs. Energy level diagrams for the three species are shown in Fig.~\ref{fig:energy_diagram}, where the transitions that were detected are indicated by blue and green arrows for PACS and SPIRE respectively. The detected transitions include H$_{2}$O lines in absorption with lower level energies $E_{\mathrm{lower}}>600$~K (H$_{2}$O $7_{07}\!\rightarrow\!6_{16}$\ and $7_{17}\!\rightarrow\!6_{06}$), in emission with upper level energies also above $600$~K (H$_{2}$O $5_{23}\!\rightarrow\!5_{14}$),and OH lines with lower level energy $E_{\mathrm{lower}}>400$~K (OH $\Pi_{1/2}\, \frac{7}{2}-\frac{5}{2}$), indicating extreme excitation similar to that in Arp~220 \citep{gon12}. Observations of highly excited H$_{2}$O over the broad spectral range provided by the combination of PACS and SPIRE give us an unique opportunity to constrain the parameters of the underlying continuum source with unprecedented accuracy. Note that the lines in the PACS range are all detected primarily in absorption while the lines detected with SPIRE are mainly observed in emission, indicating that the rotational levels are pumped at far-IR wavelengths and then relaxed by emission in the submillimeter \citep{gon10,gon14}. The analysis performed by \citet{gon14} further indicates that the absorptions are most efficiently produced close to the continuum source while the low-excitation ($E_{\mathrm{upper}}<400$~K) emission lines are formed in regions further away with less dust extinction, and as a result the combination of these wavelength ranges provides us with information on the source structure.

\begin{table*} 
\caption{H$_{2}$O, OH, and $^{18}$OH lines detected with PACS}             
\label{tab:pacslines}      
\centering                          
\begin{tabular}{l c c c c c c c}       
\hline\hline                 
Line & $\lambda_{\mathrm{rest}}$ & V$_{\mathrm{c}}\tablefootmark{a,b}$ & $\Delta V\tablefootmark{a,c}$ & $\Delta V_{\mathrm{inferred}}\tablefootmark{d}$ & Continuum\tablefootmark{e} & Peak line depth\tablefootmark{a} & Flux\tablefootmark{a} \\  
& ($\mu$m) & ($\mathrm{km\,\,s^{-1}}$) & ($\mathrm{km\,\,s^{-1}}$) & ($\mathrm{km\,\,s^{-1}}$) & (Jy) & (Jy) & ($\mathrm{Jy\,\,km\,\,s^{-1}} $) \\
\hline                       
   H$_{2}$O $4_{22}\!\rightarrow\!3_{13}$ & $57.636$ & $-25(17)$ & $319(43)$ & $247$ & $22.4$ & $1.8(0.2)$ & $-613(108)$ \\ 
   H$_{2}$O $5_{24}\!\rightarrow\!4_{13}$ & $71.067$ & $35(39)$ & $275(98)$ & $221$ & $28.7$ & $0.7(0.2)$ & $-204(93)$  \\ 
   H$_{2}$O $7_{17}\!\rightarrow\!6_{06}$ & $71.540$ & $-12(65)$ & $292(147)$ & $243$ & $29.0$ & $0.5(0.2)$ & $-144(96)$ \\
   H$_{2}$O $7_{07}\!\rightarrow\!6_{16}$ & $71.947$ & $5(33)$ & $416(103)$ & $383$ & $29.3$ & $0.8(0.2)$ & $-358(112)$  \\
   H$_{2}$O $3_{21}\!\rightarrow\!2_{12}$ & $75.381$ & $20(4)$ & $235(9)$ & $178$ & $30.9$ & $6.4(0.2)$ & $-1595(76)$ \\ 
   H$_{2}$O $4_{23}\!\rightarrow\!3_{12}$ & $78.742$ & $-1(8)$ & $239(20)$ & $190$ & $31.2$ & $3.8(0.3)$ & $-975(105)$ \\ 
   H$_{2}$O $6_{16}\!\rightarrow\!5_{05}$ & $82.031$ & $-73(26)$  &$264(73)$ & $226$ & $31.0$ & $1.7(0.3)$  &$-464(158)$  \\
   H$_{2}$O $6_{06}\!\rightarrow\!5_{15}$ & $83.284$ & $-49(25)$ & $247(71)$ & $207$ & $31.2$ & $1.3(0.3)$ & $-339(123)$  \\
   H$_{2}$O $2_{21}\!\rightarrow\!1_{10}$ & $108.073$ & $-19(10)$  &$366(26)$ & $194$ & $31.0$ & $4.0(0.2)$ & $-1541(145)$ \\
   H$_{2}$O $3_{13}\!\rightarrow\!2_{02}$\tablefootmark{f} & $138.528$ & $-4(16)$ & $286(54)$ & $109$ & $27.8$ & $2.4(1.0)$ & $-716(325)$  \\
   H$_{2}$O $3_{13}\!\rightarrow\!2_{02}$\tablefootmark{f} & $138.528$ & $266(173)$ & $508(219)$ & $434$ & $27.8$ & $-1.1(0.4)$ & $606(334)$  \\
 OH    $\Pi_{3/2}-\Pi_{3/2}\, \frac{9}{2}^--\frac{7}{2}^+$ & $ 65.132$ & $36(12)$ & $283(30)$ & $219$ & $26.6$ & $2.7(0.2)$ & $-817(111)$ \\
 OH    $\Pi_{3/2}-\Pi_{3/2}\, \frac{9}{2}^+-\frac{7}{2}^-$ & $ 65.279$ & $19(13)$ & $325(34)$ & $271$ & $26.6$ & $2.4(0.2)$ & $-833(114)$ \\ 
 OH    $\Pi_{1/2}-\Pi_{1/2}\, \frac{7}{2}-\frac{5}{2}$ & $71.197\tablefootmark{g}$ & $5(20)$ & $365(53)$ & $-$ & $28.7$ & $1.5(0.2)$ & $-596(111)$ \\ 
 OH    $\Pi_{1/2}-\Pi_{3/2}\, \frac{1}{2}^--\frac{3}{2}^+$ & $79.118$ & $-22(7)$ & $148(18)$ & $-$ & $30.8$ & $3.1(0.3)$ & $-483(76)$ \\
 OH    $\Pi_{1/2}-\Pi_{3/2}\, \frac{1}{2}^+-\frac{3}{2}^-$ & $79.181$ & $-18(7)$ & $125(19)$ & $-$ & $30.8$ & $2.9(0.4)$ & $-384(75)$ \\
 OH    $\Pi_{3/2}-\Pi_{3/2}\, \frac{7}{2}^+-\frac{5}{2}^-$ & $84.420$ & $12(4)$ & $227(9)$ & $185$  & $31.4$ & $7.2(0.2)$ & $-1742(89)$ \\
 OH    $\Pi_{3/2}-\Pi_{3/2}\, \frac{7}{2}^--\frac{5}{2}^+$ & $84.597$ & $27(4)$ & $221(9)$ & $178$ & $31.4$ & $7.3(0.2)$ & $-1712(90)$ \\
 OH    $\Pi_{3/2}-\Pi_{3/2}\, \frac{5}{2}^--\frac{3}{2}^+$ & $119.233$ & $37(5)$ & $253(11)$ & $-$ & $28.9$ & $6.8(0.3)$ & $-1842(102)$ \\
 OH    $\Pi_{3/2}-\Pi_{3/2}\, \frac{5}{2}^+-\frac{3}{2}^-$ & $119.441$ & $13(4)$ & $236(10)$ & $-$ & $28.9$ & $6.7(0.2)$ & $-1695(93)$ \\
 $^{18}$OH $\Pi_{3/2}-\Pi_{3/2}\, \frac{7}{2}^+-\frac{5}{2}^-$ & $84.947$ & $8(9)$ & $99(23)$ & $-$ & $32.6$ & $1.8(0.4)$ & $-191(59)$ \\
 $^{18}$OH $\Pi_{3/2}-\Pi_{3/2}\, \frac{7}{2}^--\frac{5}{2}^+$ & $85.123$ & $-39(20)$ & $201(48)$ & $-$ & $32.6$ & $1.3(0.3)$ & $-273(86)$ \\
 $^{18}$OH    $\Pi_{3/2}-\Pi_{3/2}\, \frac{5}{2}^+-\frac{3}{2}^-$ & $120.171$ & $-36(33)$ & $254(82)$ & $-$ & $29.3$ & $0.8(0.2)$ & $-227(96)$ \\

\hline                                   
\end{tabular}
\tablefoot{
\tablefoottext{a}{Values from Gaussian fits to the lines, numbers in parenthesis indicate 1$\sigma$ uncertainties from these fits.}
\tablefoottext{b}{Velocity shift of line center relative to $z = 0.012999$.}
\tablefoottext{c}{FWHM of lines.}
\tablefoottext{d}{Inferred velocity width is based on the instrument resolution assuming a Gaussian profile; this is not listed for doublets that are not well separated.}
\tablefoottext{e}{Value of the fitted baseline at the line center.}
\tablefoottext{f}{The P-Cygni profile has been fitted using a combination of one absorption and one emission component.}
\tablefoottext{g}{The two $\Lambda-$components are (nearly) blended into a single spectral feature.}
}
\end{table*}

\begin{table} 
\caption{H$_{2}$O and H$_{2}^{18}$O lines detected with SPIRE.}             
\label{tab:spirelines}      
\centering                          
\begin{tabular}{l c c c c}        
\hline\hline                 
Line & E$_{\mathrm{upper}}$ & $\lambda_{\mathrm{rest}}$ & Cont.\tablefootmark{a} & Flux\tablefootmark{b} \\  
& (K) &($\mu$m) & (Jy) & ($\mathrm{Jy\,\,km\,\,s^{-1}} $) \\
\hline                        
   H$_{2}$O $1_{11}\!\rightarrow\!0_{00}$\tablefootmark{c} & $53$ & $269.272$ & $6.9$ & $83.5(35.5)$  \\
   H$_{2}$O $1_{11}\!\rightarrow\!0_{00}$\tablefootmark{c} & $53$ & $269.272$ & $6.9$ & $-146.8(35.5)$  \\
   H$_{2}$O $2_{02}\!\rightarrow\!1_{11}$ & $101$ & $303.456$ & $5.0$ & $904.1(53.3)$ \\
   H$_{2}$O $2_{11}\!\rightarrow\!2_{02}$ & $137$ & $398.643$ & $2.2$ & $667.5(70.1)$ \\
   H$_{2}$O $2_{20}\!\rightarrow\!2_{11}$ & $196$ & $243.974$ & $9.0$ & $507.6(60.6)$  \\
   H$_{2}$O $3_{12}\!\rightarrow\!2_{21}$\tablefootmark{d} & $249$ & $259.982$ & $7.7$ & $-$  \\
   H$_{2}$O $3_{12}\!\rightarrow\!3_{03}$ & $249$ & $273.193$ & $6.7$ & $812.2(80.2)$ \\
   H$_{2}$O $3_{21}\!\rightarrow\!3_{12}$ & $305$ & $257.795$ & $7.8$ & $1225.6(56.1)$  \\
   H$_{2}$O $4_{22}\!\rightarrow\!4_{13}$ & $454$ & $248.247$ & $8.6$ & $617.9(57.5)$  \\
   H$_{2}$O $5_{23}\!\rightarrow\!5_{14}$ & $642$ & $212.526$ & $12.6$ & $200.8(37.6)$  \\
  H$_{2}^{18}$O $2_{20}\!\rightarrow\!2_{11}$\tablefootmark{e} & $194$ & $250.034$ & $8.6$ & $171.5(44.3)$ \\
  H$_{2}^{18}$O $3_{21}\!\rightarrow\!3_{12}$ & $303$ & $263.738$ & $7.4$ & $328.1(64.4)$  \\

\hline                                   
\end{tabular}
\tablefoot{
 \tablefoottext{a}{Value of the fitted baseline at the line center.}
\tablefoottext{b}{Values from fits of a convolution of a sinc function and a Gaussian function to the lines, numbers in parenthesis indicate 1$\sigma$ uncertainties from these fits.}
 \tablefoottext{c}{The tentative P-Cygni profile in this line has been fitted using a combination of one absorption and one emission component. During the fitting, the frequencies for these components were held fixed at the positions of maximum emission and absorption in the feature.}
 \tablefoottext{d}{Blended with CO $J=10-9$, the total flux in the line is $1135.5(79.2)$~$\mathrm{Jy\,\,km\,\,s^{-1}}$.}
 \tablefoottext{e}{Likely to be affected by the sinc profile of the nearby H$_{2}$O $4_{22}\!\rightarrow\!4_{13}$ line.}
}
\end{table}

\subsection{H$_{2}$O}
Ten transitions of H$_{2}$O with lower level energies up to $\sim 650$~K, were detected with PACS as summarized in Table \ref{tab:pacslines}. The high excitation in the lines indicates that they might be radiatively excited \citep[e.g.,][]{gon04,gon08,gon12} and can be used to constrain the properties of the underlying continuum source. The spectral scans of the H$_{2}$O lines are shown in Fig. \ref{fig:h2o}. The absorption is not as strong as in NGC 4418, but is comparable to that of Arp~220 \citep{gon12}. In addition, nine transitions with upper level energies up to $\sim 650$~K were detected primarily in emission with SPIRE as summarized in Table \ref{tab:spirelines}. The spectral line energy distribution of the submillimeter lines is shown in Fig. \ref{fig:h2osled}. As mentioned earlier, the emission lines with upper level energies $E_{\mathrm{upper}}<400$~K have been shown to trace regions with low dust extinction, while the two detected emission lines with higher $E_{\mathrm{upper}}$ are found to form in very warm ($>80$~K) regions with higher dust extinction \citep{gon14}. Most of the lines are uncontaminated by other species, but the H$_{2}$O $3_{12}\!\rightarrow\!2_{21}$\ line is blended with the CO $J=10-9$ transition and is not used in the analysis of the excitation.

\subsubsection{P-Cygni profile in H$_{2}$O $3_{13}\!\rightarrow\!2_{02}$}\label{sec:pcygni}
Although some lines appear shifted relative to the systemic velocity we find no systematic velocity shift, with some lines slightly shifted to the red and others to the blue. Also, within $2\sigma$, all but two H$_{2}$O lines peak at the galaxy redshift. The $3_{13}\!\rightarrow\!2_{02}$\ line at $138$~$\mu$m does however exhibit an apparent P-Cygni profile, possibly indicating an outflow. Signatures of this are not seen in any of the other PACS lines but out of our detected H$_{2}$O absorption lines, $3_{13}\!\rightarrow\!2_{02}$\ is the one with the longest wavelength. It is thus the line that is least affected by dust extinction and it is possible that the redshifted emission traces an outflow in the outer component on the far side of the galaxy that cannot be seen in the lines at shorter wavelengths, including the nearby $2_{21}\!\rightarrow\!1_{10}$\ line at $108$~$\mu$m, due to extinction. A possible explanation for the difference in character of the two nearby lines, is that the $2_{21}\!\rightarrow\!1_{10}$\ line, which is lower in energy, is also tracing the infall seen in the [O I] $63$~$\mu$m line (see Sect. \ref{sec:oi}), in which case the redshifted absorption and emission could cancel. In fact, the profile of the $2_{21}\!\rightarrow\!1_{10}$\ line is slightly asymmetric, with less absorption on the redshifted side. The only H$_{2}$O line detected in absorption in SPIRE (H$_{2}$O $1_{11}\!\rightarrow\!0_{00}$, see insert in Fig. \ref{fig:full_spire}) tentatively exhibits a profile similar to the H$_{2}$O $3_{13}\!\rightarrow\!2_{02}$ line, with absorption at zero velocity and a redshifted emission component.

\subsubsection{H$_{2}^{18}$O}
Two transitions in the isotopologue H$_{2}^{18}$O were also detected with SPIRE: $2_{20}\!\rightarrow\!2_{11}$ and $3_{21}\!\rightarrow\!3_{12}$\ with upper level energies of $194$ and $303$~K, respectively. While the $3_{21}\!\rightarrow\!3_{12}$ line seems to be uncontaminated by other species, the $2_{20}\!\rightarrow\!2_{11}$\ is likely to be affected by the sinc profile of the nearby H$_{2}$O $4_{22}\!\rightarrow\!4_{13}$ line.

   \begin{figure}  
    \centering
    \includegraphics[width=8.0cm]{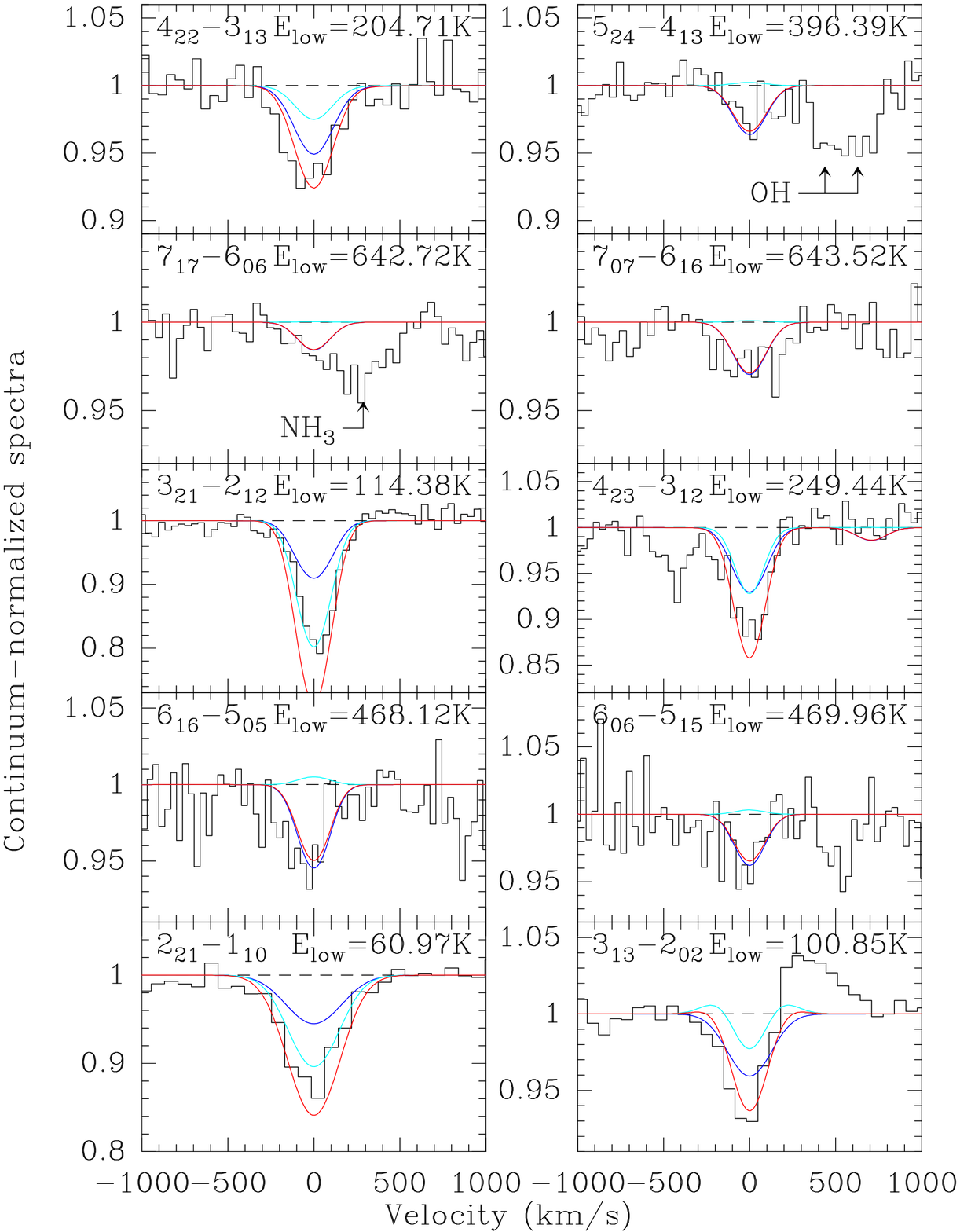}
    \caption{ H$_{2}$O lines observed in Zw~049.057 with PACS. The black histograms are the observed, continuum-normalized, spectra. Model results from Sect. \ref{sec:models} are also included. The blue curve denotes the model for $C_{\mathrm{core}}$, which was constrained using the high-lying H$_{2}$O lines detected with PACS. The light-blue curve denotes the model for $C_{\mathrm{outer}}$, which was constrained using the H$_{2}$O lines detected with SPIRE. The red curve denotes the sum of these two models.}
    \label{fig:h2o}
   \end{figure}

   \begin{figure} 
    \centering
    \includegraphics[width=8.0cm]{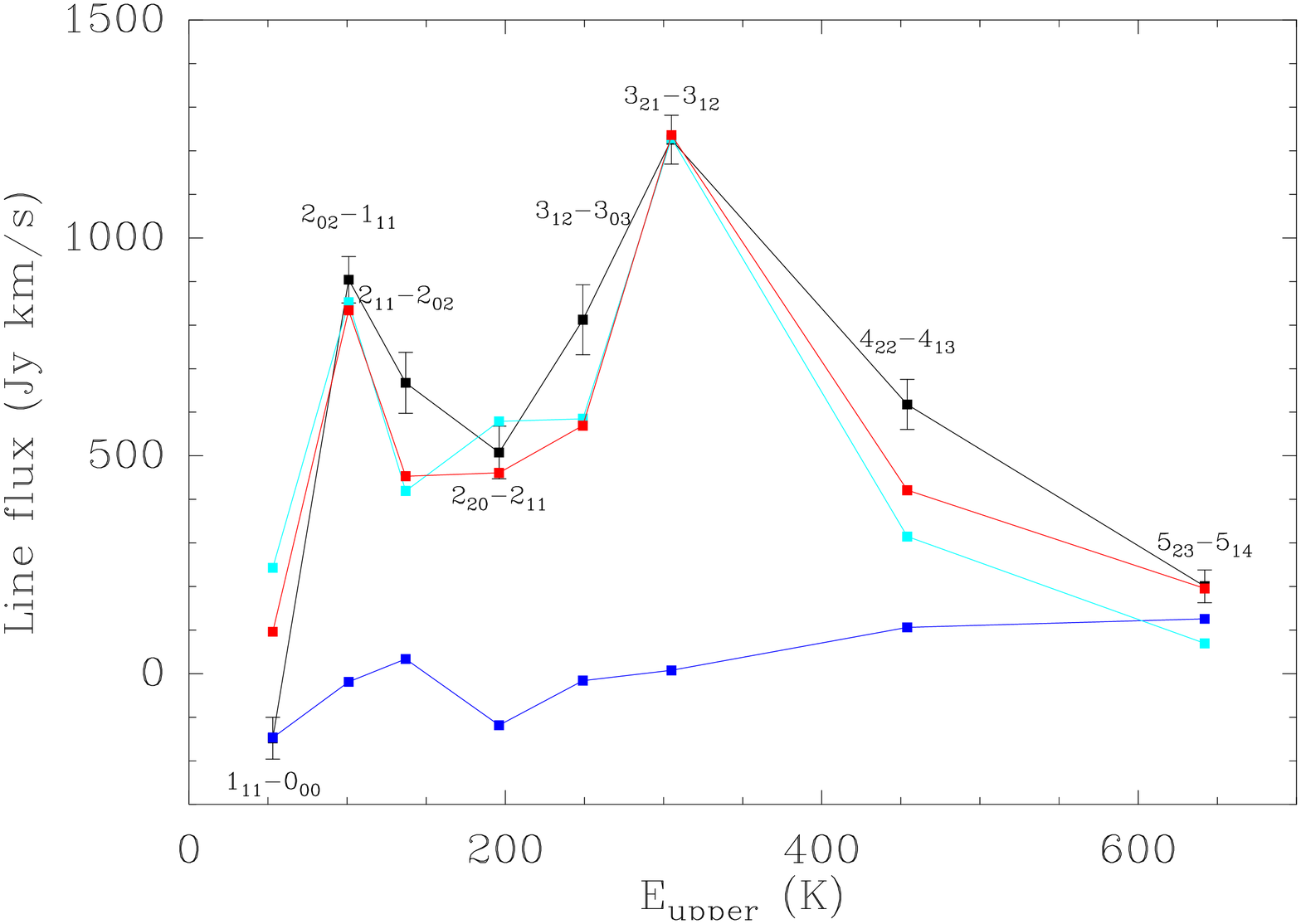}
    \caption{Spectral line energy distribution of the H$_{2}$O lines detected with SPIRE, the black curve represents the data. For the $1_{11}\!\rightarrow\!0_{00}$ line, only the absorption part is included. The line with upper energy level $249$~K is $3_{12}\!\rightarrow\!3_{03}$. Model predictions are included, the blue and light-blue curves show the contribution from $C_{\mathrm{core}}$ and $C_{\mathrm{outer}}$, respectively. The red curve indicates the sum of the fluxes predicted by the two models.}  
    \label{fig:h2osled}
   \end{figure}
\subsection{OH} 
\begin{figure} 
    \centering
    \includegraphics[width=8.0cm]{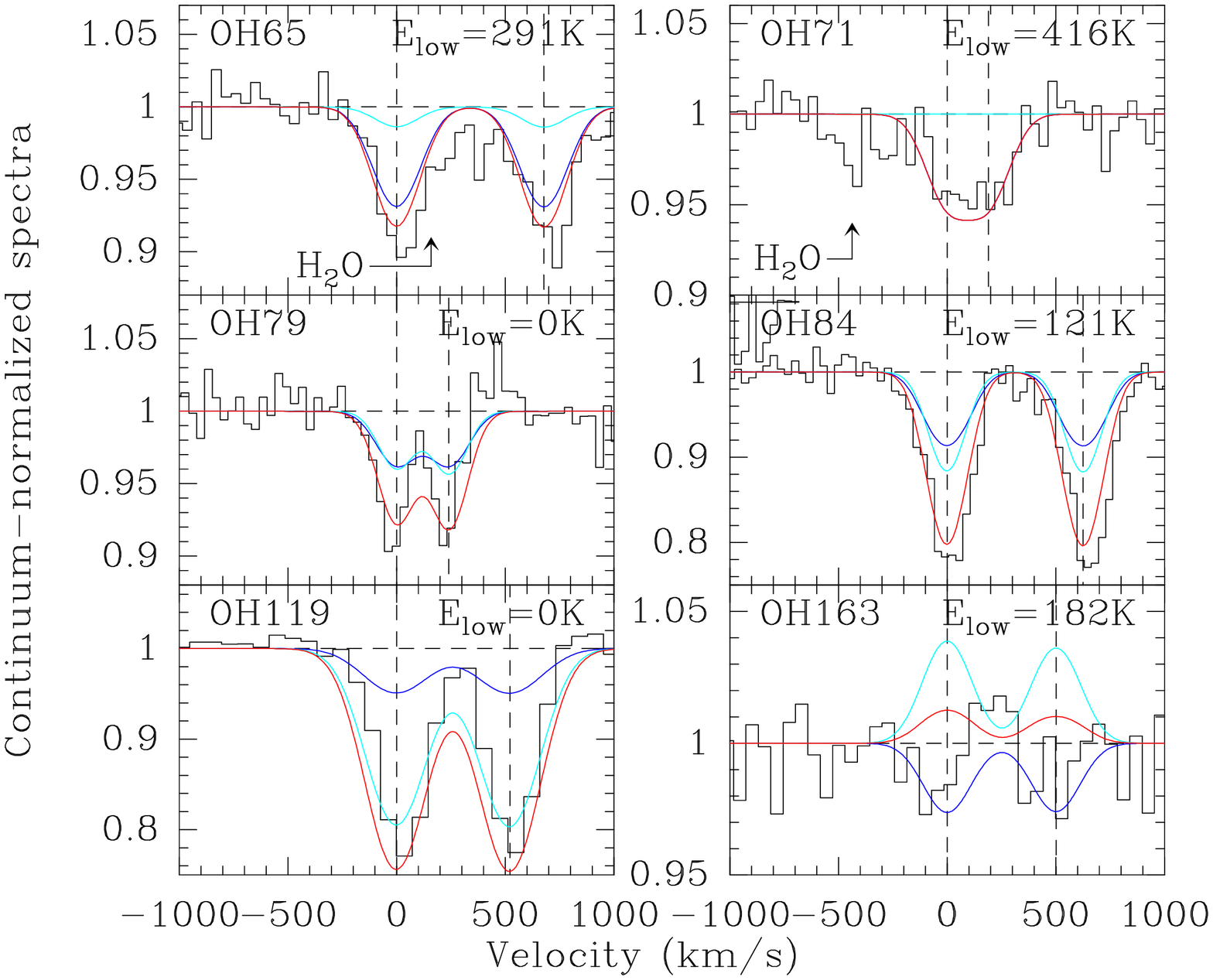}
    \caption{OH lines observed in Zw~049.057 with PACS. The black histograms are the observed, continuum-normalized, spectra. Model results from Sect. \ref{sec:models} are also included. The blue curve denotes the model for $C_{\mathrm{core}}$, which was first constrained using the high-lying H$_{2}$O lines detected with PACS, but with an added OH column that was varied in order to also fit the high-lying OH lines. The light-blue curve denotes the model for $C_{\mathrm{outer}}$, which was first constrained using the H$_{2}$O lines detected with SPIRE, but with an added OH column that was varied in order to also account for the residual flux in the low-lying OH lines. The red curve denotes the sum of these two models.} 
    \label{fig:oh}
   \end{figure}

Five OH doublets with lower level energies up to $>400$~K were detected with PACS, they are summarized in Table \ref{tab:pacslines}. The spectral scans of these lines are shown in Fig. \ref{fig:oh}, together with that of the undetected $\Pi_{1/2} \, 3/2\rightarrow1/2$ doublet at 163 $\mu$m. Again, the absorptions are not as strong as in NGC 4418, but in the most excited lines they are almost comparable to those in Arp 220 \citep{gon12}.

The doublet at 65 $\mu$m might have weak contamination by H$_{2}$O $6_{25}\!\rightarrow\!5_{14}$ in the blue $\Lambda$-component and the 84 $\mu$m doublet might be weakly contaminated by NH$_{3}$~\hbox{$(6,4)a\!\rightarrow\!(5,4)s$} and \hbox{$(6,5)a\!\rightarrow\!(5,5)s$} in the red $\Lambda$-component. 

\subsubsection{$^{18}$OH}
\begin{figure}
    \centering
    \includegraphics[width=8.0cm]{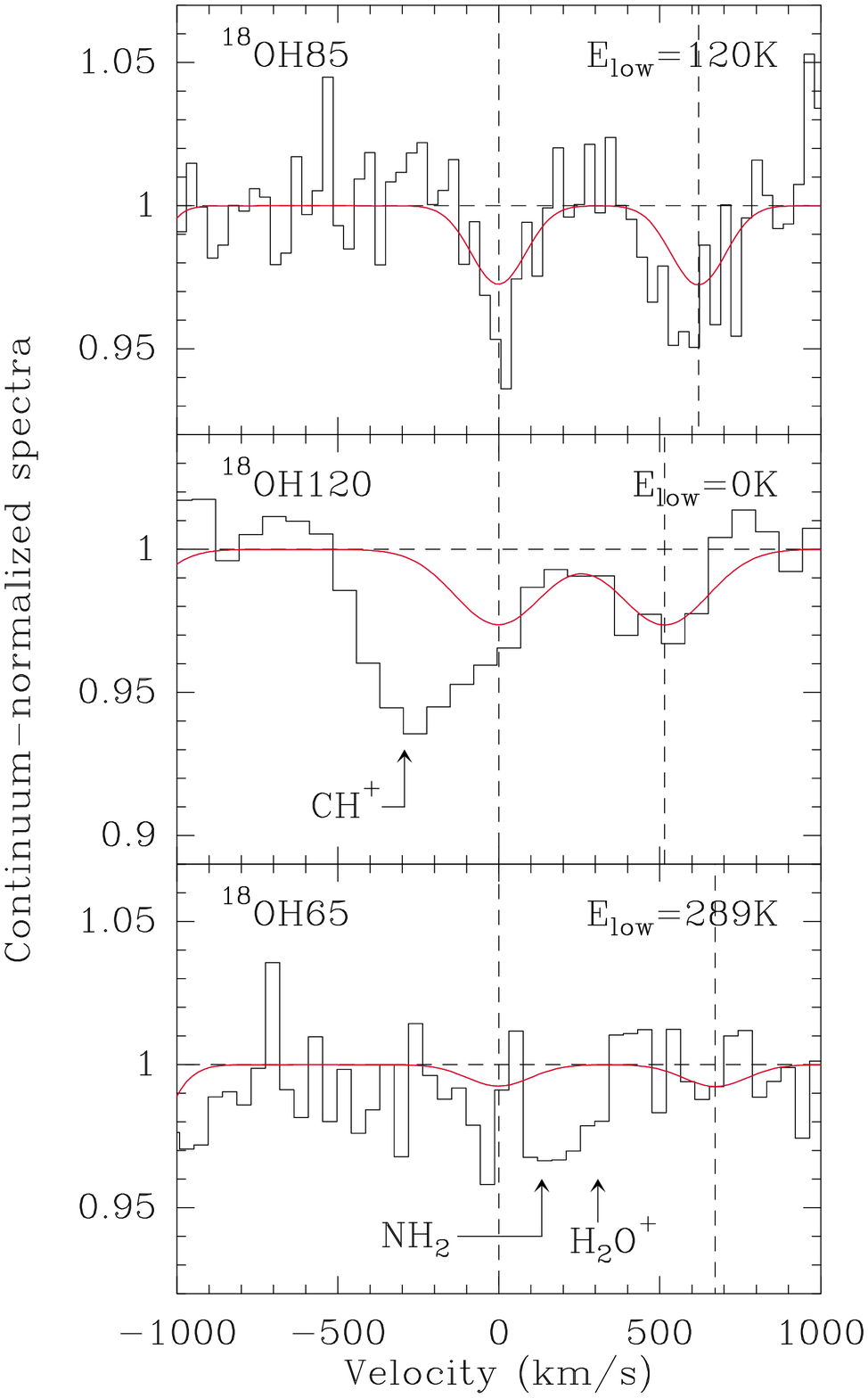}
    \caption{$^{18}$OH lines observed in Zw~049.057 with PACS. The black histograms are the observed, continuum-normalized, spectra. Model results from Sect. \ref{sec:models} are also included. The red curve denotes the model for $C_{\mathrm{core}}$, which was first constrained using the high-lying H$_{2}$O lines detected with PACS, but with an added $^{18}$OH column that was varied in order to fit the $^{18}$OH lines.}  
    \label{fig:18oh}
   \end{figure}

Two doublets of the isotopologue $^{18}$OH were also detected: $\Pi_{1/2} \, 7/2\rightarrow5/2$ at 85 $\mu$m and $\Pi_{3/2} \, 5/2\rightarrow3/2$ at 120 $\mu$m. The spectral scans of these lines are shown in Fig. \ref{fig:18oh}, together with that of the undetected $\Pi_{3/2} \, 9/2\rightarrow7/2$ doublet at 65 $\mu$m. The blue $\Lambda$-component of the 120 $\mu$m doublet is strongly contaminated by CH$^+$ 3-2. Although the absorption in the main isotopologue generally seems to be weaker than observed in NGC 4418 by \citet{gon12}, the $^{18}$OH absorption is stronger in Zw~049.057. This is the same situation as \citet{gon12} found in Arp 220, likely indicating enhancement of $^{18}$O in Zw~049.057 as was inferred for Arp 220.  

\subsection{[O I] $63$~$\mu$m}\label{sec:oi}
\begin{figure} 
    \centering
    \includegraphics[width=8.0cm]{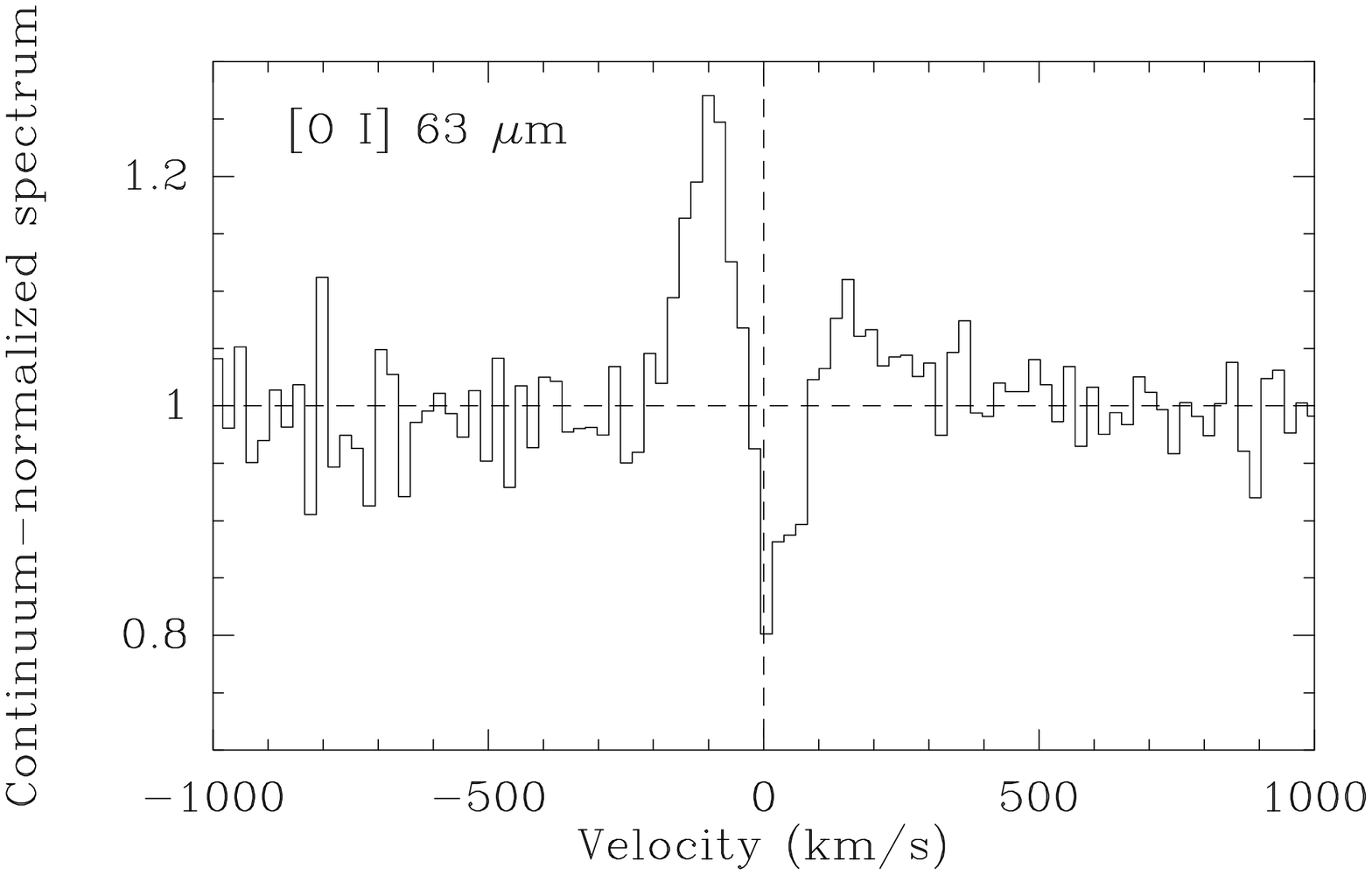}
    \caption{Line profile of the [O I] 63~$\mu$m line detected with PACS. The reversed P-Cygni profile in this line is a typical signature of infalling gas.}  
    \label{fig:oi}
   \end{figure}

The spectral scan of the [O I] $63$~$\mu$m line is shown in Fig. \ref{fig:oi}, note the prominent reversed P-Cygni profile, a typical signature of infalling gas. We do not attempt to model this line, but it provides important information about the kinematics on scales larger than the molecular components in our models. 

\subsection{Full SPIRE FTS spectrum} \label{sec:full_spire}
\begin{figure*} 
  \centering
   \includegraphics[width=18.0cm]{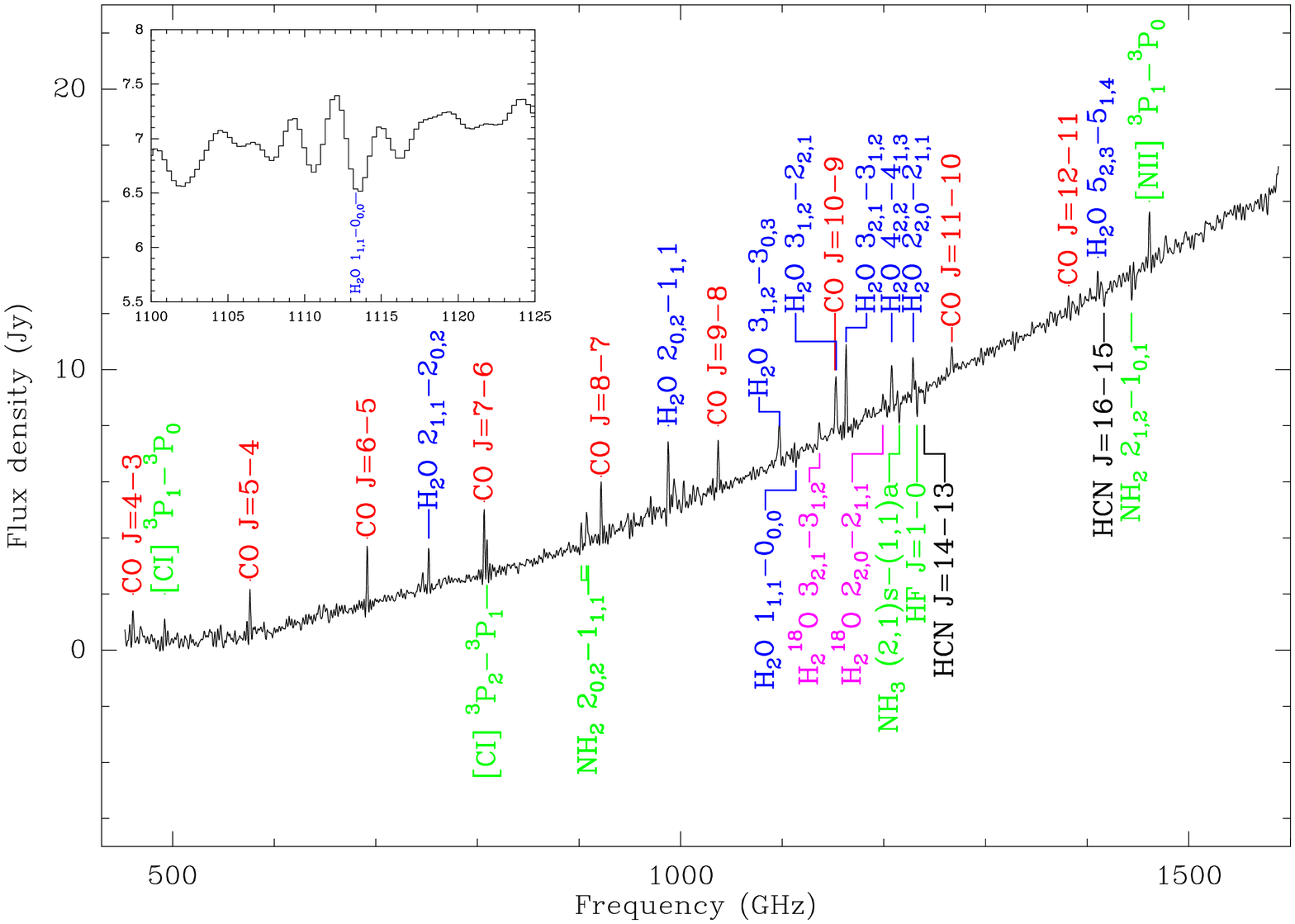}
   \caption{Full SPIRE FTS spectrum of Zw 049.057. Line identifications for $\geq 3\sigma$ detections are given in red for CO lines, in blue for H$_{2}$O, in magenta for H$_{2}^{18}$O, in black for HCN, and in green for the remaining lines. The insert shows a zoomed-in view of the H$_{2}$O $1_{11}\!\rightarrow\!0_{00}$ line.}   
   \label{fig:full_spire}
\end{figure*}

The full SPIRE FTS spectrum of Zw~049.057 is shown in Fig.~\ref{fig:full_spire}. Because of the excellent match in the overlap region of the two spectrometer bands ($\sim 950-1020$~GHz), they were simply averaged in this region. The spectrum shows at least $30$ lines detected at a $3\sigma$ level or higher. Most of the CO ladder is detected with nine lines from CO $J = 4-3$ to $J = 12-11$\citep{gre14,ros15}. In addition, nine rotational lines of H$_{2}$O are detected together with two rotational lines of H$_{2}^{18}$O. The two [C\,{\sc i}] fine structure lines and the [N\,{\sc ii}] fine structure line as well as two rotational transitions of HCN are also detected. Finally, one rotational transition of NH$_{3}$ as well as one of HF, and at least three transitions of NH$_{2}$ are detected. There are hints of absorption accompanied with redshifted emission at the positions of the two OH$^{+}$ lines at $972$ and $1033$~GHz, but the noise in this region is too high for an unambiguous detection. Much stronger OH$^{+}$ lines have been observed in emission in Mrk~231 \citep{van10} and with P-Cygni profiles in Arp~220 \citep{ran11}. The relative strength of the H$_{2}$O lines with respect to the CO lines is greater in Zw~049.057 than in Mrk~231 and comparable to that seen in Arp~220. The lines are detected superposed on a continuum that drops towards the long wavelength side, and represents the Rayleigh-Jeans tail of the dust emission in Zw~049.057.

\section{Models}
\label{sec:models}
As shown by the data presented in Sect. \ref{sec:observations} the far-IR PACS spectra of Zw~049.057 are dominated by molecular absorption, with only a few emission features, while the opposite is true for the submillimeter SPIRE spectrum. Because the absorption and emission lines of H$_{2}$O require different physical conditions to form \citep{gon14}, the galaxy must contain regions of different ISM parameters to produce these rich line spectra and their associated dust continua. We attempt to model these regions using the smallest possible number of parameterized components. In doing so, we use the spherically symmetric radiative transfer code described in \citet{gon97,gon99}, including line overlaps between between the $\Lambda-$components of the OH doublets in calculations for OH. Dust emission is simulated by using a mixture of silicate and amorphous carbon grains with optical constants from \citet{dra85} and \citet{pre93}, the adopted mass absorption coefficient as a function of wavelength is described in \citet{gon14}. Rates for collisional excitation of H$_{2}$O are taken from \citet{dub09} and \citet{dan11} while the rates for OH were taken from \citet{off94}. Note that the modeled regions are spatially unresolved by the \emph{Herschel} beams. 

We find that at least two components are needed to fit the line absorption/emission and the continuum at wavelengths $\lesssim 60$~$\mu$m. Depicted for the continuum in Fig. \ref{fig:continuum}: a compact warm component (blue curve, hereafter $C_{\mathrm{core}}$) to account for the high-lying H$_{2}$O lines and a larger cool component (light blue curve, $C_{\mathrm{outer}}$) to account for the low-lying OH lines as well as most of the flux in the submillimeter emission lines. The two components that account for all H$_{2}$O and OH absorption/emission lines cannot, however, reproduce the observed continuum at $\lambda>60$~$\mu$m, so that an additional cold ($T_{\mathrm{d}}=30$~K) continuum component is added in Fig. \ref{fig:continuum} (green curve) to fit the whole SED. This extra component has no associated H$_{2}$O/OH spectrum, which is expected given the relatively high $T_{\mathrm{d}}$ required to excite these species. The parameters inferred for the two H$_{2}$O/OH components are listed in Tables \ref{tab:cont} and \ref{tab:lines}. In addition to these components, an even hotter and more compact component would be needed to fit the emission at wavelengths shorter than $\sim10$~$\mu$m as was also found for NGC 4418 and Arp 220 by \citet{gon12}.

\subsection{Outline of the modeling}
In the following models we have assumed a standard gas-to-dust ratio of $\sim100$ as found in the central regions of other LIRGs by \citet{wil08}. The high excitation in the molecular lines detected in absorption towards the nucleus cannot be accounted for by collisions alone, strongly suggesting that the dominant excitation mechanism is absorption of radiation emitted by warm dust as in the nuclear regions of Mrk 231, Arp 220, and NGC 4418 \citep{gon04,gon08,gon10,gon12}. Our models indeed indicate that collisional excitation in the core component described below is negligible in comparison with radiative excitation for reasonable densities and gas temperatures ($n_{\mathrm{H_{2}}}\lesssim5\times10^{6}$~cm$^{-3}$, $T_{\mathrm{gas}}\lesssim 500$~K), and thus our model results are insensitive to $T_{\mathrm{gas}}$. In the outer component, where the radiation field is weaker, collisional excitation is expected to have some impact on the low-lying submillimeter H$_{2}$O lines (Sect. \ref{sec:outer} below), but the excitation of the high-lying lines is dominated by radiative pumping. The excitation thus provides important clues about the properties of the continuum source and about the column densities of H$_{2}$O and OH. In all models the molecules are mixed with the dust, but as in Arp 220 \citep{gon12} the models indicate that most of the absorption is formed in the outer layers of the far-IR source where the continuum optical depth is $\lesssim 1$. The molecular column densities are therefore given for the outermost region in which $\tau_{50}=1$, where $\tau_{50}$ is the dust opacity at $50$~$\mu$m as measured from the observer. In case a model over predicts the flux density at $25$~$\mu$m, it is attenuated by a foreground absorbing shell parameterized by its dust opacity at $25$~$\mu$m. Evidence for such an absorbing layer comes from the strong silicate absorption at $9.7$~$\mu$m \citep{per10} as well as from the absorption in the [O I] $63$~$\mu$m line (see Fig. \ref{fig:oi}). The total attenuated luminosity is however never allowed to exceed $10^{11.27}$ L$_{\sun}$, which is the total IR luminosity of the galaxy \citep{san03}. Details of the modeling are discussed in Sects. \ref{sec:core} and \ref{sec:outer}.

In the models for the $C_{\mathrm{core}}$ component, where collisional excitation is negligible, the line ratios depend on the following free model parameters: the dust temperature ($T_{\mathrm{d}}$), the column of dust (parameterized by the continuum optical depth at 100 $\mu$m, $\tau_{100}$), the column density of H$_{2}$O per unit of $\tau_{50}$ ($N_{\mathrm{H_{2}O}}/\tau_{50})$, and the velocity dispersion ($\Delta V$, which is fixed, and derived from the observed line widths). In addition, absolute line and continuum fluxes depend on the size of the component (parameterized by the radius $R$, see Tables \ref{tab:cont} and \ref{tab:lines}). As pointed out above, results of the models for the $C_{\mathrm{outer}}$ component are also sensitive to the collisional excitation, and thus also depend on $n_{\mathrm{H_{2}}}$ and $T_{\mathrm{gas}}$. For these latter parameters, we adopt the results by \citet[][$T_{\mathrm{gas}}=200$~K and $n_{\mathrm{H_{2}}}=10^5$~cm$^{-3}$]{man13a,man13}.

Our fitting strategy was the following: we developed a grid of models for H$_{2}$O in both the $C_{\mathrm{core}}$ and $C_{\mathrm{outer}}$ components (some illustrative results are shown in Figs. \ref{fig:lineratiospacs} and \ref{fig:lineratiosspire}) by varying  $T_{\mathrm{d}}$, $\tau_{100}$, and $N_{\mathrm{H_{2}O}}/\tau_{50}$. Since both the $C_{\mathrm{core}}$ and $C_{\mathrm{outer}}$ components generally contribute to the flux of a given H$_{2}$O line, we fitted all far-IR and submillimeter H$_{2}$O lines simultaneously by combining different parameters for $C_{\mathrm{core}}$ and $C_{\mathrm{outer}}$. Once the best model fit for $C_{\mathrm{core}}$ and $C_{\mathrm{outer}}$ was obtained from all H$_{2}$O lines, the same components were applied to OH by varying only the OH column in both components.

\subsection{The core component}\label{sec:core}
\begin{figure} 
   \centering
   \includegraphics[width=8.0cm]{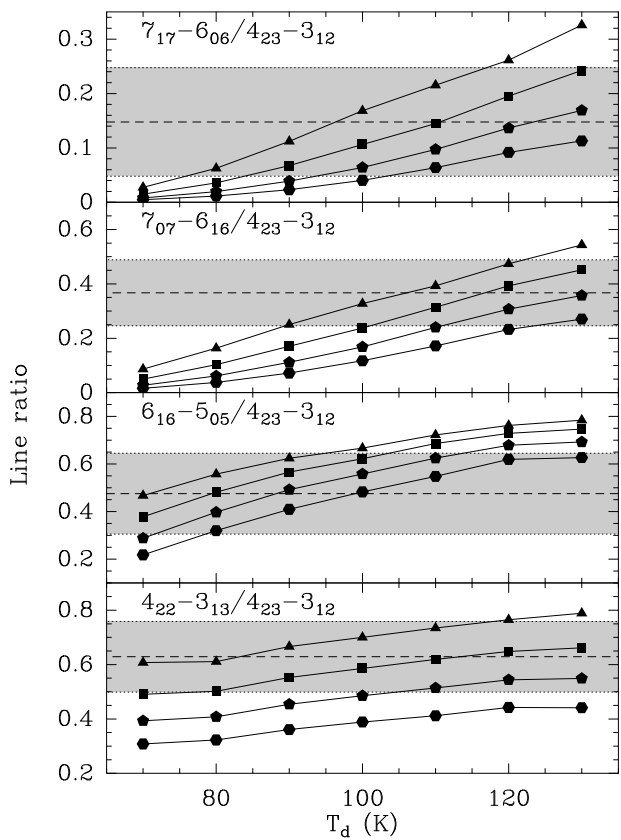}
   \caption{Modeled H$_{2}$O line ratios in $C_{\mathrm{core}}$ as a function of dust temperature. Triangles, squares, pentagons, and circles show results for H$_{2}$O columns per unit of $\tau_{50}$ of $2 \times 10^{18}$, $1 \times 10^{18}$, $5 \times 10^{17}$, and $2.5 \times 10^{17}$ cm$^{-2}$, respectively. Dashed lines indicate the observed ratios and the dotted lines are their $1\sigma$ uncertainties.}  
   \label{fig:lineratiospacs}
  \end{figure}

The best constraints on the dust temperature ($T_{\mathrm{d}}$) and column density ($N_{\mathrm{H_{2}O}}$) come from the high-lying lines of H$_{2}$O, which are expected to be radiatively excited. In Fig. \ref{fig:lineratiospacs} we compare the observed and modeled fluxes of some of the highly excited lines, normalized to the flux in the $4_{23}\!\rightarrow\!3_{12}$ line. This line is chosen as the normalization because it is detected with high signal-to-noise and its lower level is high enough in energy for the line to remain relatively unaffected by extended low-excitation H$_{2}$O. The modeled ratios are plotted as a function of $T_{\mathrm{d}}$ and $N_{\mathrm{H_{2}O}}/\tau_{50}$. We ran the models in Fig. \ref{fig:lineratiospacs} both with and without collisions included and found that the line ratios are unaffected by collisional excitation for $n_{\mathrm{H_{2}}}\lesssim5\times10^{6}$~cm$^{-3}$ and $T_{\mathrm{gas}}\lesssim 500$~K. For higher densities and gas temperatures the submillimeter H$_{2}$O $1_{11}\!\rightarrow\!0_{00}$ line will go into strong emission, which is not seen in the SPIRE spectrum. Line broadening is modeled by microturbulence with $v_{\mathrm{tur}}=60$ km~s$^{-1}$. 

The relative fluxes of the high-lying lines can be well fitted with dust temperatures between $T_{\mathrm{d}}=90$ and $130$~K and column densities per $\tau_{50}$ between $N_{\mathrm{H_{2}O}}/\tau_{50}=2 \times 10^{18}$ and $5 \times 10^{17}$~cm$^{-2}$, although not all combinations of $T_{\mathrm{d}}$ and $N_{\mathrm{H_{2}O}}/\tau_{50}$ in this range produce a good fit. A lower dust temperature could also be used to fit the absorptions if a higher column density is used, but then the submillimeter emission lines $4_{22}\!\rightarrow\!4_{13}$ and $5_{23}\!\rightarrow\!5_{14}$ would be over predicted. In order to fit the absolute line fluxes the modeled region must have a radius between $32$ and $12$~pc. At these temperatures and radii, a foreground absorbing shell with an opacity between $\tau_{25}=2.1$ and $\tau_{25}=2.4$ is required in order to fit the flux density at $25$~$\mu$m.

The dust opacity is best constrained by the high-lying submillimeter emission lines. With the column densities and temperatures required to fit the absorptions in the far-IR the emission line with the highest upper level energy, $5_{23}\!\rightarrow\!5_{14}$, would be over predicted for $\tau_{100}\lesssim5$ and under predicted for $\tau_{100}\gtrsim15$. In order to fit the low-lying submillimeter lines of H$_{2}$O an additional, less optically thick component is required, which is discussed in Sect. \ref{sec:outer}.  

To fit the high-lying OH lines we used the model parameters derived for H$_{2}$O and varied the column density of OH to fit the observed line fluxes. We find the best fit for OH/H$_{2}$O ratios of $0.4-0.8$, higher OH column densities over predict the $\Pi_{1/2}-\Pi_{1/2}\, \frac{7}{2}-\frac{5}{2}$ doublet at $71$~$\mu$m while lower column densities do not produce enough absorption in the $\Pi_{3/2}-\Pi_{3/2}\, \frac{9}{2}-\frac{7}{2}$ doublet at $65$~$\mu$m. To fit the low-lying OH lines another component, which is discussed in Sect. \ref{sec:outer}, is required. The doublets of the isotopologue $^{18}$OH require high column densities in order to obtain a good fit. We find that an $^{16}$OH/$^{18}$OH ratio of $50-100$ is required to reproduce the absorption doublet at $120$~$\mu$m while the one at $85$~$\mu$m is slightly under predicted. 

The contribution of the core component to the total model is shown in blue in Figs. \ref{fig:continuum} and \ref{fig:h2o}-\ref{fig:18oh}. In this reference model we have used $T_{\mathrm{d}}=120$~K, $N_{\mathrm{H_{2}O}}/\tau_{50}=1 \times 10^{18}$~cm$^{-2}$, $\tau_{100}=10$, $R=14$~pc, OH/H$_{2}$O$=0.75$, and $^{16}$OH/$^{18}$OH$=75$.

\subsection{The outer component}\label{sec:outer}
\begin{figure} 
   \centering
   \includegraphics[width=8.0cm]{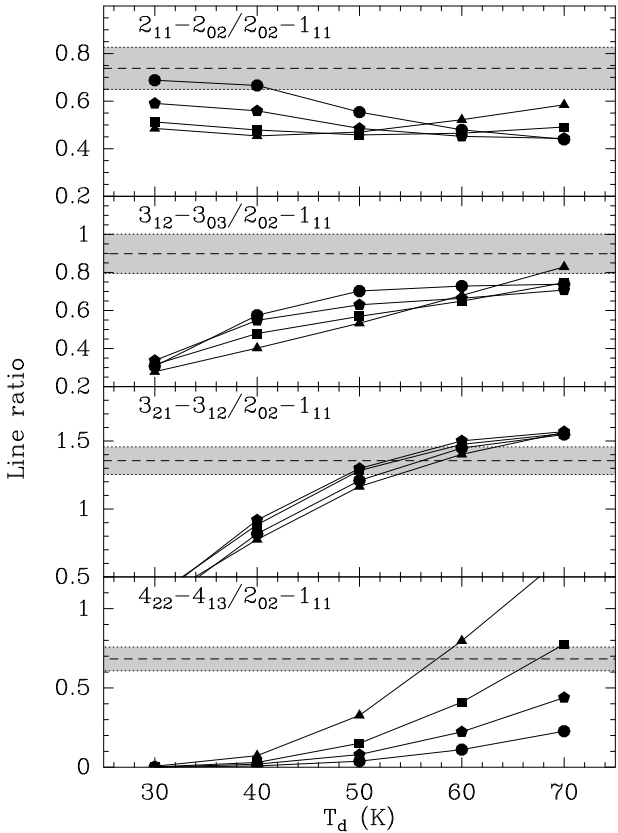}
   \caption{Modeled H$_{2}$O line ratios in $C_{\mathrm{outer}}$ as a function of dust temperature. Triangles, squares, pentagons, and circles show results for H$_{2}$O columns per unit $\tau_{50}$ of $8 \times 10^{17}$, $4 \times 10^{17}$, $2 \times 10^{17}$, and $1 \times 10^{17}$, respectively. In all models $\tau_{100}=1$. Dashed lines indicate the observed ratios and the dotted lines are their $1\sigma$ uncertainties. The contribution from $C_{\mathrm{core}}$ is not included in this figure, but it is expected to slightly raise the $2_{11}\!\rightarrow\!2_{02}/2_{02}\!\rightarrow\!1_{11}$ and  $4_{22}\!\rightarrow\!4_{13}/2_{02}\!\rightarrow\!1_{11}$ ratios. The overall best fit, including the core contribution, is shown in Fig. \ref{fig:h2osled}}  
   \label{fig:lineratiosspire}
  \end{figure}

To adequately fit the emission in the submillimeter H$_{2}$O lines and the absorption in the low-lying OH lines another, less excited, component is required. This component also accounts for the remaining flux in the lower-lying H$_{2}$O absorption lines, which are under predicted by the core component alone. In Fig. \ref{fig:lineratiosspire} we compare the observed and modeled fluxes of some of the submillimeter lines, normalized to the flux of the $2_{02}\!\rightarrow\!1_{11}$ line. The core component is expected to provide extra emission in the $2_{11}\!\rightarrow\!2_{02}$ and $4_{22}\!\rightarrow\!4_{13}$ lines. Some of the submillimeter lines, for example H$_{2}$O $4_{22}\!\rightarrow\!4_{13}$, are however still not reproduced, probably indicating a transition region between the two components where the emission in these lines is still significant. In this component we find that the ratios of the lowest lying lines are significantly altered by collisional excitation, in the models shown here we have adopted a gas temperature $T_{\mathrm{g}}=200$~K and a density $n_{\mathrm{H_{2}}}=10^{5}$~cm$^{-3}$ as derived by \citet{man13a,man13} based on observations of formaldehyde and ammonia. A potential caveat is that they did not consider IR pumping when estimating their temperatures. The main effect of increasing the density is to strengthen the low-lying H$_{2}$O $1_{11}\!\rightarrow\!0_{00}$\ and $2_{02}\!\rightarrow\!1_{11}$\ lines at the expense of the H$_{2}$O $3_{21}\!\rightarrow\!3_{12}$\ line. With a higher density the two lower lines would be over predicted while a lower density produces too much emission in the H$_{2}$O $3_{21}\!\rightarrow\!3_{12}$\ line.

In order to reproduce the line ratios, we find that a dust temperature between $T_{\mathrm{d}}=45$ and $60$~K and column densities per $\tau_{50}$ between $N_{\mathrm{H_{2}O}}/\tau_{50}=2 \times 10^{17}$ and $8 \times 10^{17}$~cm$^{-2}$ are required. The line ratios are also sensitive to the dust opacity, we find that the ratios can be reproduced with $\tau_{100}$ from $0.5$ to $2$. In order to reproduce the absolute fluxes the component must have a radius of $90$ to $50$~pc. The line widths are still best reproduced using $v_{\mathrm{tur}}=60$ km~s$^{-1}$.

The submillimeter H$_{2}^{18}$O $3_{21}\!\rightarrow\!3_{12}$ line is mainly formed in the outer component, requiring an abundance ratio $^{16}$O/$^{18}$O$\lesssim 50$ in water. With this ratio for OH, the $^{18}$OH $120$~$\mu$m doublet would be overestimated by a factor $\sim 2$, but the $^{18}$OH $85$~$\mu$m doublet would be better reproduced than in our reference model (Fig. \ref{fig:18oh}). We thus favor a range $^{16}$O/$^{18}$O$=50-100$, with enhanced $^{18}$O abundance in both the core and outer components.

The OH lines are again fitted with the model derived for the H$_{2}$O excitation. In this component a OH/H$_{2}$O ratio of $0.1$ to $0.4$ is required to fit the low-lying OH lines. With higher OH column densities the undetected doublet at $163$~$\mu$m is over predicted.

The contribution of the outer component to the total model is shown in light blue in Figs. \ref{fig:continuum} and \ref{fig:h2o}-\ref{fig:18oh}. In this reference model we have used $T_{\mathrm{d}}=55$~K, $N_{\mathrm{H_{2}O}}/\tau_{50}=4 \times 10^{17}$~cm$^{-2}$, $\tau_{100}=1$, $R=60$~pc, and OH/H$_{2}$O$=0.2$. 

   \begin{table*} 
      \caption[]{Parameters of the continuum models.}
         \label{tab:cont}
          \centering
          \begin{tabular}{lcccccccc}   
            \hline\hline 
            \noalign{\smallskip}
            C\tablefootmark{a} & Radius\tablefootmark{b} & $T_{\mathrm{d}}$\tablefootmark{b} & $\tau_{100}$\tablefootmark{b} & $N_{\mathrm{H_{2}}}\tablefootmark{c}$ & $M\tablefootmark{d}$ & $L\tablefootmark{e}$ & $\tau_{\mathrm{25,fgr}}\tablefootmark{f}$ & $L_{\mathrm{att}}\tablefootmark{g}$ \\ 
              & (pc) & (K) & at 100 $\mu$m & (cm$^{-2}$) & ($10^8$ M$_{\sun}$) & (L$_{\sun}$) & & (L$_{\sun}$) \\ 
            \noalign{\smallskip}
            \hline
         $C_{\mathrm{core}}$ & $32-12$ & $90-130$ & $5-15$ & $(3-10)\times10^{24}$ & $0.3-7$ & $(0.7-1.2)\times10^{11}$ & $2.1-2.4$ & $(2-6)\times10^{10}$ \\ 
         $C_{\mathrm{outer}}$ & $90-50$  & $45-60$ &  $0.5-2$ & $(0.3-1.3)\times10^{24}$ & $0.5-7$ & $(4-9)\times10^{10}$ & $0$ & $(4-9)\times10^{10}$ \\ 
            \noalign{\smallskip}
               \noalign{\smallskip}
            \hline
         \end{tabular} 
\tablefoot{
\tablefoottext{a}{Component.}
\tablefoottext{b}{Independent parameter.}
\tablefoottext{c}{Column density of H$_{2}$, calculated assuming a mass-absorption coefficient of 44.4 cm$^2$~g$^{-1}$ at 100 $\mu$m and a gas-to-dust mass ratio of 100.}
\tablefoottext{d}{Estimated mass, assuming spherical symmetry.}
\tablefoottext{e}{Unattenuated luminosity of the component.}
\tablefoottext{f}{Foreground opacity at 25 $\mu$m.}
\tablefoottext{g}{Attenuated luminosity of the component.}
}
   \end{table*}

  \begin{table*} 
      \caption[]{Derived H$_{2}$O column densities and abundances, and column density
        ratios}
         \label{tab:lines}
         \centering
          \begin{tabular}{lccccc}   
            \hline\hline 
            \noalign{\smallskip}
            C\tablefootmark{a} & $N_{\mathrm{H_2O}}/\tau_{50}\tablefootmark{b,c}$ & $\chi_{\mathrm{H_2O}}\tablefootmark{d}$ & OH/H$_{2}$O\tablefootmark{b,e} &  $^{16}$O/$^{18}$O\tablefootmark{b,e} \\
            &  (cm$^{-2}$)  &  &   &  \\
            \noalign{\smallskip}
            \hline
            $C_{\mathrm{core}}$  & $(0.5-2)\times10^{18}$  & $(2.5-10)\times10^{-6}$ & $0.4-0.8$ & $50-100$ \\
            $C_{\mathrm{outer}}$ & $(2-8)\times10^{17}$  & $(1-4)\times10^{-6}$ & $0.1-0.4$ & $50-100$  \\
            \noalign{\smallskip}
               \noalign{\smallskip}
            \hline
         \end{tabular}
\tablefoot{
\tablefoottext{a}{Component.}
\tablefoottext{b}{Independent parameter.}
\tablefoottext{c}{Column densities per unit of dust opacity at 50~$\mu$m, $\tau_{50}$.}
\tablefoottext{d}{Estimated H$_{2}$O abundance relative to H nuclei.}
\tablefoottext{e}{Column density ratios.}
}
   \end{table*}

\subsection{Model uncertainties}
We have ignored all inner structure in the modeled components by adopting uniform densities and dust temperatures throughout the models. The approximation of spherical symmetry is also clearly inaccurate, the NICMOS image of Zw~049.057, for example, exhibits a linear structure extending for hundreds of parsecs \citep{sco00}. For the absorption lines, geometry should however be secondary as the important property is the projected surface of the underlying optically thick far-IR source. Our models should nevertheless be considered as very simplified versions of the real structures.

The dust temperature in $C_{\mathrm{core}}$ is well constrained as high temperatures are required for the radiatively excited high-lying lines to form. In $C_{\mathrm{outer}}$ we expect that collisional excitation is also important and $T_{\mathrm{d}}$ is here less certain than in $C_{\mathrm{core}}$. Due to dust extinction, the absorption lines are sensitive to the molecular column densities only in the outer layers of the core where $\tau_{50}\lesssim1$. In these layers they are however fairly certain and present robust lower limits to the total column densities of the source. The conversion of $\tau_{100}$ to a column density of H$_{2}$ depends on the adopted gas-to-dust ratio as well as the dust properties. This uncertainty is propagated to the molecular abundances, which of course depend on the column of H$_{2}$. Abundance ratios between different species are more certain as the various abundances are calculated in the same way. 

\section{Discussion}
\label{sec:discussion}
\begin{sidewaystable*}
\caption{Physical conditions in other objects for comparison}             
\label{tab:comparison}      
\centering                          
\begin{tabular}{l c c c c c c c c c }       
\hline\hline                 
Object & Class & $N_{\mathrm{H_{2}}}$ & $\chi_{\mathrm{H_2O}}$ & $\chi_{\mathrm{OH}}$ & $R$ & $T_{\mathrm{d}}$ & $M$ & Ref. \\  
& & (cm$^{-2}$) & & & (pc) & (K) & ($10^8$ M$_{\sun}$) & \\
\hline                       
  APM 08279+5255 & HyLIRG & $-$  & $6 \times 10^{-7}\tablefootmark{a}$ & $-$ & $100$ & $-$ & $-$ & 1,2 \\ 
  Mrk 231$_{\mathrm{warm}}$& ULIRG & $-$ & $8.0 \times 10^{-7}$ & $-$ & $120$ & $95$ & $5.9$ & 3 \\ 
  Mrk 231$_{\mathrm{extended}}$ & ULIRG & $-$ &  $-$ & $-$ & $610$ & $41$ & $77$ & 3 \\ 
  Arp 220$_{\mathrm{west}}$ & ULIRG & $(3-6) \times 10^{25}$ &  $(0.4-3) \times 10^{-5}$ & $(0.16-3) \times 10^{-5}$ & $75-40$ & $90-130$ & $55-30$ & 4 \\ 
  Arp 220$_{\mathrm{east}}$ & ULIRG & $1.3 \times 10^{25}$ & $1.6 \times 10^{-6}$ & $\sim 1.6 \times 10^{-6}$ & $54$ & $87$ & $13$ & 4 \\ 
  Arp 220$_{\mathrm{extended}}$ & ULIRG & $8.0 \times 10^{23}$ & $8 \times 10^{-8}$ & $\sim 2.4 \times 10^{-7}$ & $325$ & $90-40$ & $30$ & 4 \\ 
  NGC 4418$_{\mathrm{core}}$ & LIRG & $(0.5-1.0) \times 10^{25}$ & $(1-3) \times 10^{-5}$ & $(0.3-1.8) \times 10^{-5}$ & $10$ & $140-150$ & $0.16-0.33$ & 4 \\ 
  NGC 4418$_{\mathrm{warm}}$ & LIRG & $2.4 \times 10^{24}$ & $(1-5) \times 10^{-7}$ & $(0.4-2) \times 10^{-7}$ & $15.5$ & $110$ & $0.2$ &  4 \\ 
  NGC 4418$_{\mathrm{extended}}$ & LIRG & $(1.0-1.3) \times 10^{23}$ &  $4 \times 10^{-8}$ & $(2-4) \times 10^{-7}$ & $100$ & $90-30$ & $0.18$ &  4 \\ 
  Zw~049.057$_{\mathrm{core}}$\tablefootmark{b} & LIRG & $(3-10) \times 10^{24}$ & $(2.5-10) \times 10^{-6}$ & $(1-8) \times 10^{-6}$ & $32-12$ & $90-130$ & $0.3-7$ & 5 \\ 
  Zw~049.057$_{\mathrm{outer}}$\tablefootmark{b} & LIRG & $(0.3-1.3) \times 10^{24}$ & $(1-4) \times 10^{-6}$ & $(0.1-1.6) \times 10^{-6}$ & $90-50$ & $45-60$ & $0.5-7$ & 5 \\ 
\hline\hline
Object & Class & $N_{\mathrm{H_{2}}}$ & $\chi_{\mathrm{H_2O}}$ & $\chi_{\mathrm{OH}}$ & $R$ & $T_{\mathrm{d}}$ & $M$ & Ref. \\  
& & (cm$^{-2}$) & & & (AU) & (K) & ($10^8$ M$_{\sun}$) & \\
\hline                       
 Sgr B2 & Denser regions & $-$ & $\mathrm{few} \times 10^{-7}$ & $-$ & $-$ & $-$ & $-$ & 6 \\
 Sgr B2 & Envelope & $-$ & $(1-2)\times 10^{-5}$ & $(2-5)\times 10^{-6}$ & $-$ & $-$ & $-$ &  6,7 \\
 Orion KL & Hot core & $-$  & $1.0 \times 10^{-5}$ & $-$ & $-$ & $-$ & $-$ & 8 \\
 Orion KL & Plateau & $-$ & $7.4 \times 10^{-5}$ & $(0.5-1.0) \times 10^{-6}$ & $-$ & $-$ & $-$ & 8,9 \\
 Orion-IRc2 & Shocked low-velocity gas & $-$ &  $(2-5)\times 10^{-4}$ & $(6.2-15.6)\times 10^{-6}$ & $-$ & $-$ & $-$ & 10 \\ 
 NGC 6334 I & Hot core & $-$ & $\sim 10^{-6}$ & $-$ & $-$ & $-$ & $-$ & 11 \\
 NGC 1333 IRAS2A & Hot core & $-$ & $>2 \times 10^{-5}$ & $-$ & $\sim 100$ & $-$ & $-$ &  12 \\
 W43-MM1 & Hot core & $-$ & $1.4 \times 10^{-4}$ & $-$ & $\sim 3500$ & $-$ & $-$ & 13 \\
 L1157-B1 & Warm shocked region & $1.2 \times 10^{20}$ & $(0.7-2) \times 10^{-6}$ & $-$ & $1250$ & $-$ & $-$ & 14 \\
 L1157-B1 & Hot shocked region & $3.3 \times 10^{20}$ & $(1.2-3.6) \times 10^{-4}$ & $-$ & $250-625$ & $-$ & $-$ & 14 \\
\hline                                   
\end{tabular}
\tablefoot{
\tablefoottext{a}{The value from \citet{van11} has been used.}\\
}
\tablebib{
(1) \citet{bra11}; (2) \citet{van11}; (3) \citet{gon10}; (4) \citet{gon12}; (5) This work; (6) \citet{cer06}; (7) \citet{goi02}; (8) \citet{mel10}; (9) \citet{goi06}; (10) \citet{wri00}; (11) \citet{emp13}; (12) \citet{vis13}; (13) \citet{her12}; (14) \citet{bus14}
}
\end{sidewaystable*}

\subsection{The dust and gas components of Zw~049.057}
As described in Sect. \ref{sec:models} above we can fit the spectral line data with two components whose properties are summarized in Tables \ref{tab:cont} and \ref{tab:lines}. For comparison, Table \ref{tab:comparison} contains a summary of physical conditions in some other galaxies as well as in some Galactic regions. 

Compared to giant molecular clouds (GMCs) in the Galaxy, $C_{\mathrm{core}}$ is comparable in size but significantly hotter as well as more massive and luminous \citep{sco89}. Although cooler and less dense, it shows strong similarity to the compact hot core found in the LIRG NGC~4418 \citep{gon12}. The physical conditions are also very similar to those in the western nucleus of the ULIRG Arp~220, which is a few times as big as $C_{\mathrm{core}}$ \citep{gon12}.  

The outer component is larger than a typical GMC and is (again) significantly more massive and luminous, its dust temperature is however comparable to the peak temperature found in H II-region GMCs \citep{sco89}. An interesting comparison is with the Galactic Center - where a mass of a few times $10^{7}$~M$_{\sun}$ of molecular gas is distributed on a scale of $200-500$ pc \citep{sco87,cox89,mor96,dah98,fer07}. Thus the nuclear region of Zw~049.057 appears to be dramatically different from the center of our own galaxy, with an order of magnitude more molecular gas concentrated on scales comparable to a large GMC. 

\subsection{Abundances and column densities toward Zw~049.057}
All abundances derived for Zw~049.057 are summarized in Table \ref{tab:lines}. For comparison, a summary of abundance estimates in other galaxies and selected Galactic regions is tabulated in Table \ref{tab:comparison}. 

\subsubsection{H$_{2}$O abundances}
We find high abundances of H$_{2}$O in $C_{\mathrm{core}}$ - lower in $C_{\mathrm{outer}}$. In the compact inner component the H$_{2}$O abundance is similar to those found in the core component of NGC~4418 and the western nucleus of Arp~220 \citep{gon12}. There are no regions in the Galaxy where an H$_{2}$O abundance this high has been found over comparable scales. On smaller scales abundances comparable to, and higher than, that in $C_{\mathrm{core}}$ have been found in hot cores and shocked regions \citep[e.g.,][]{mel10,emp13,vis13,bus14}. Comparing it to more extreme objects further out in the Universe we see that Zw~049.057 has a H$_{2}$O abundance exceeding that of the H$_{2}$O rich $z=3.91$ quasar APM~08279+5255 \citep{bra11}. 

In the outer component, $C_{\mathrm{outer}}$, we find H$_{2}$O abundances on the order of $10^{-6}$. This is similar to the abundances found by \citet{gon10} and \citet{gon12} in the warm component of Mrk~231 and the eastern nucleus of Arp~220, respectively. In the Galaxy, similar H$_{2}$O abundances have been found, for example, in the denser regions of Sgr~B2 by \citet{cer06} and the warm shocked region of L1157-B1 by \citet{bus14}.

\subsubsection{Why is there so much water in Zw 049.057?}
The high H$_{2}$O abundance in the core component is a robust result so we will focus our discussion here. There are three main scenarios which can explain H$_{2}$O abundances this high: ion-neutral reactions starting with cosmic-ray (or X-ray) ionization of H$_{2}$ \citep{van13}, neutral-neutral reactions in warm (T~$\gtrsim 250$~K) gas \citep{neu95,cec96,van13}, and sublimation of H$_{2}$O from dust grains as the dust temperature rises above $\sim 100$~K \citep{san90,fra01,van13}. If the high H$_{2}$O abundance is due to the first of these processes it can persist as long as the temperature of the gas remains high enough. If it is due to sublimation from dust grains however, the abundance will eventually decrease unless there is an efficient formation process to counter H$_{2}$O destruction by, for example, ion-neutral reactions.

The model of $C_{\mathrm{core}}$ indicates that the dust temperature is slightly higher than that required for rapid sublimation to occur while the gas temperature ($189\pm57$~K) as estimated by \citet{man13} is somewhat lower than that needed for neutral-neutral reactions to be effective. Also, we know that an additional, hotter, component is needed to model the continuum at wavelengths $<10$~$\mu$m. Formation of H$_{2}$O via ion-neutral reactions may contribute to the high abundances, but the branching ratios in the dissociative recombination of H$_{3}$O$^{+}$ favor OH \citep{her90,jen00,neu02} while we infer more H$_{2}$O than OH in both components. 

In view of the uncertainties above, we are unable to determine which one of the processes is responsible for the high H$_{2}$O abundance, and a combination is most likely. One way to test the grain chemistry would be to search for CH$_{3}$OH, which seems to be mainly formed through hydrogenation of CO on grain surfaces \citep{wir11}.

\subsubsection{OH abundances}
The OH abundance found in $C_{\mathrm{core}}$ is, like the H$_{2}$O abundance, similar to those found in the core of NGC~4418 and the western nucleus of Arp~220 by \citet{gon12}. We also find that the abundance of OH seems to vary less than the H$_{2}$O abundance between $C_{\mathrm{core}}$ and $C_{\mathrm{outer}}$. For Galactic sources we found fewer estimates of OH abundances, but there are regions around Sgr~B2 and the Orion nebula that have abundances that are comparable to those in $C_{\mathrm{core}}$ and $C_{\mathrm{outer}}$ \citep{goi02,goi06}.

\subsection{A Compton-thick nucleus in Zw~049.057?}
A source is, by definition, Compton-thick if the column density of the obscuring matter is equal to or greater than the inverse of the Thomson cross-section, i.e., if N$_{H}\geq \sigma_{\mathrm{T}}^{-1} \simeq 1.5 \times 10^{24}$~cm$^{-2}$. The column density that we find in $C_{\mathrm{core}}$ is thus enough to make the nucleus Compton-thick and a buried X-ray source would be difficult to detect. In order to penetrate the obscuring layers and probe the nuclear energy source in Zw~049.057 observations in hard X-rays or even soft gamma rays are therefore needed.

The highly obscured nature of Zw~049.057 makes it difficult to unambiguously determine the source of its nuclear luminosity. Below we investigate two possible scenarios. 

\subsubsection{Obscured AGN}
A possible AGN in Zw~049.057 was indicated by \citet{baa06} based on radio observations and their nuclear classification code. With the extreme column density indicated by our model a hidden AGN is also allowed despite the relatively weak X-ray luminosity observed by \citet{leh10} with \emph{Chandra}. A way to keep a high covering factor for the obscuring matter has been proposed by \citet{fab98}. They suggested a model in which a nuclear starburst both feeds the central black hole and hides it by injecting the energy needed to keep the obscuring matter in a highly turbulent and space-covering state. In this scenario the nucleus stays in the obscured state until the accreting black hole has grown to a certain limit were the accretion rate approaches the Eddington limit and the obscuring matter is blown away from the nucleus.

If all of the luminosity in $C_{\mathrm{core}}$ comes from an accreting black hole we can derive a lower limit to its mass of $\sim 3.8-5.6 \times 10^{6}$ M$_{\sun}$ by assuming that it radiates at its Eddington limit. In reality, due to the high optical depth of the dust continuum the limit on the mass can actually be up to $500$ times higher \citep{fab08}. 

The similarity with the western nucleus of Arp 220, suggested by some authors to harbor a hidden AGN, warrants a detailed comparison. A central concentration of hard X-ray emission in the western nucleus of Arp~220 was found by \citet{cle02} using \emph{Chandra} X-ray imaging. Several possible scenarios for this emission are offered, including a weak AGN contributing less than 1\% to the bolometric luminosity of the galaxy. They also state that they cannot rule out the possibility that they are just seeing a small fraction of the emission from an obscured AGN and refer to the hard X-ray \emph{BeppoSAX} observations by \citet{iwa01} who found that a column higher than $\sim 10^{25}$~cm$^{-2}$ is required if a large fraction of the bolometric luminosity comes from an AGN. Based partly on the very high surface brightness of $5 \times 10^{14}$~L$_{\sun}$~kpc$^{-2}$ on a scale of $35$~pc found in their millimeter interferometer observation, \citet{dow07} concluded that the source of the luminosity in the western nucleus of Arp~220 can only be a black hole accretion disk. They also estimated that the total proton column density towards the western nucleus is $\sim 1.3 \times 10^{25}$~cm$^{-2}$, enough to hide the hard X-rays from an energetically important AGN. In addition, \citet{tak05} found absorption lines of NH$_{3}$ with full widths at zero intensity of up to $1800$~km~s$^{-1}$ possibly due to rotating gas in an AGN. \citet{sak08}, however, do not exclude that the luminosity is primarily powered by a young starburst equivalent to hundreds of super star clusters.

Zw~049.057, on the other hand, was undetected in hard X-rays in the \emph{Chandra} observations by \citet{leh10}. However, as already stated, our models indicate that the column density towards the center of Zw~049.057 may be as high as $10^{25}$~cm$^{-2}$, comparable to that in Arp~220. Observations by \citet{man13} reveals that Zw~049.057 also has NH$_{3}$ absorption lines. These are however not nearly as wide as those seen in Arp~220. High spatial resolution observations at millimeter wavelengths, like those of \citet{dow07}, has not been carried out for Zw~049.057. Doing this would be one way to truly test how much it actually resembles the western nucleus of Arp~220, and possibly also to determine if it is powered by an AGN or a central starburst.

\subsubsection {Starburst}
\label{sec:starburst}
Another possible candidate for the nuclear power source is a compact starburst. \citet{vei95} classified Zw~049.057 as an H\,{\sc ii} galaxy based on optical emission line ratios in the nucleus of the galaxy. Based on mid-IR observations \citep{sti13,ina13} it is also located in the starburst region in the diagnostic diagram by \citet{spo07} as well as in mid-IR emission line diagrams \citep{mel14}.

The core component has a luminosity of $0.7-1.2 \times10^{11}$~L$_{\sun}$ and a radius of $12-32$~pc. To put these numbers into context we can use Sgr~B2(M) with its luminosity of $6.3 \times 10^{6}$ L$_{\sun}$ \citep{gol92} and radius of $\sim 0.6$ pc \citep{rol11} as a template. To generate the luminosity of $C_{\mathrm{core}}$, $\sim(1-2)\times10^{4}$ Sgr~B2(M)-like sources would be required to be packed into a volume between $0.13$ and $1.25$ times that of $C_{\mathrm{core}}$.

The surface brightness of the core component is very high, $0.4-1.5 \times 10^{14}$~L$_{\sun}$~kpc$^{-2}$. This is up to one order of magnitude higher than the limiting surface brightness in nuclear starbursts around AGN seen by \citet{dav07}. For dust temperatures below $200$~K this apparent limit for starbursts can be explained by a theoretical model in which stellar radiation pressure provides the majority of pressure support in the gas disk \citep{tho05}. For hot starburst (T$_{\mathrm{d}} > 200$~K) this limit does not apply and the surface brightness can be higher than $\sim 10^{13}$~L$_{\sun}$~kpc$^{-2}$ \citep{and11}. The core component is not hot enough for this to happen, but our models do not reproduce the continuum at wavelengths $<10$~$\mu$m and it is possible that a hot nuclear component, which we do not model, has a dust temperature $> 200$~K. In any case, \citet{and11} further state that the conditions necessary to enter this regime may only be attained in the parsec-scale star formation surrounding AGN.

Taking only the gas mass into account, our model for $C_{\mathrm{core}}$ implies a luminosity-to-mass ratio of $\sim200-2000$~L$_{\sun}$~M$_{\sun}^{-1}$, largely compatible with the limit of $500-1000$~L$_{\sun}$~M$_{\sun}^{-1}$ for a dust-embedded starburst supported by radiation pressure \citep{sco03,tho05}. 

\subsection{In- and outflowing gas}
We find a strong inverse P-Cygni profile in the [O~I] $63$~$\mu$m line, indicating inflowing gas. No obvious evidence for such inflow is however seen in the molecular lines so the infall might take place on scales larger than the $\sim 100$~pc molecular structure. There is however a possibility that the H$_{2}$O $2_{21}\!\rightarrow\!1_{10}$ line is actually tracing both infalling and outflowing gas, see the discussion in Sect. \ref{sec:pcygni}. The estimated 60 micron source size is $\sim 100$~pc, so the red-shifted gas motions have to occur on these (projected) scales. 

A possible source of infall could be tidal debris or interactions where a galaxy accretes gas from a companion (as is the case for NGC4418, for example; Klöckner, priv. comm.). There are several galaxies in the same field as Zw~049.057, but they seem to belong to a background cluster and are not companions of Zw~049.057. Another possibility is gas streaming in from a cooling flow similar to those seen in elliptical galaxies \citep{bre04}. Bar instabilities that acts to drive gas into the nucleus is a possible scenario but it seems unlikely since it would require rather a precise geometry.

We also find a normal P-Cygni profile in the H$_{2}$O $3_{13}\!\rightarrow\!2_{02}$\ line, indicative of outflowing gas. We do not see any outflow signatures in the other molecular lines however, and if the feature is real it probably traces gas outflowing on the far side of the galaxy as discussed in Sect. \ref{sec:pcygni}. Compared to the fastest molecular outflows seen in Herschel studies of LIRGs \citep[e.g.,][]{fis10,stu11,vei13,gon14a} the potential outflow velocities here are moderate, with the redshifted velocity component extending to $\sim500$~km~s$^{-1}$. 

\subsection{The $^{16}$O/$^{18}$O isotopic ratio}
$^{18}$O is a so called secondary nuclide, produced by He burning \citep{mae83} in massive stars and may enrich the interstellar medium (ISM) by winds and/or supernova ejecta. Hence the $^{16}$O/$^{18}$O isotopic abundance ratio may reflect the stellar initial mass function (IMF) as well as the age of a burst of star formation. Low $^{16}$O/$^{18}$O abundance ratios are taken to indicate the enrichment of the ISM by massive stars in a starburst. \citet{gon12} use this ratio to place LIRGs/ULIRGs in an evolutionary scenario where the dust enshrouded LIRG NGC~4418 (with an estimated $^{16}$O/$^{18}$O ratio of $500$) is placed in an early evolutionary phase, the ULIRG Arp~220 (with an estimated  $^{16}$O/$^{18}$O ratio of $70$) is more evolved according to this scheme. The ULIRG quasar Mrk 231 has an extreme $^{16}$O/$^{18}$O ratio of $\lesssim30$ \citep{fis10,gon14a} and hence has the most evolved starburst of the three. A caveat here is that it is assumed that all three galaxies have starbursts of a similar IMF with the same impact and ISM mixing. Note that a large $^{16}$O/$^{18}$O ratio may also be a signature of a very evolved starburst where the $^{16}$O has caught up with the $^{18}$O enrichment.

We estimate a $^{16}$O/$^{18}$O abundance in Zw~049.057 that is similar to or slightly higher than that in Arp 220, and this would therefore suggest that its starburst is in an evolutionary stage between NGC~4418 and Arp~220. This is difficult to validate since the nuclei are so deeply obscured that standard tracers of starburst evolution do not work. A possibility is to use the q-factor (i.e., the ratio between radio and far-IR intensity \citep{hel85}) -- but this is aggravated by the fact that Mrk~231 is a quasar so its radio emission is dominated by the jet. However, if we take the $^{16}$O/$^{18}$O ratios as tracing starburst evolution and compare the ''younger'' NGC~4418 with the ''older'' Zw~049.057 we find interesting similarities between them. Both have strong infall signatures in the [O~I] $63$~$\mu$m line suggesting that the central region of the galaxies are being replenished with new material that can feed the activity. NGC~4418 is interacting while this is not clear for Zw~049.057. Both galaxies are early type spirals where Zw~049.057 is showing evidence also of star formation on somewhat larger scales than NGC~4418. No outflow signatures are found in the gas of NGC~4418 (even though there are some hints in the dust \citep{sak13}) while Zw~049.057 is showing tentative outflow signatures.


\section{Conclusions}
\label{sec:conclusions}
The following results were obtained from the observations of Zw~049.057 presented in this paper:
   \begin{itemize}
      \item PACS spectroscopy reveals absorption in highly excited states of H$_{2}$O and OH as well as in their $^{18}$O isotopologues. In total, ten H$_{2}$O lines and five OH doublets with lower level energies up to $E_{\mathrm{lower}}~600$~K and $E_{\mathrm{lower}}~400$~K, respectively, were detected in absorption. Two $^{18}$OH doublets were also detected in absorption.
      \item The [O I] $63 \mu$m line exhibits an inverse P-Cygni profile indicative of infalling gas with a velocity of $~30$ km~s$^{-1}$. A P-Cygni profile in the H$_{2}$O $3_{13}\!\rightarrow\!2_{02}$\ line with the redshifted emission extending to $~500$~km~s$^{-1}$ was also detected.
      \item SPIRE spectroscopy reveals a submillimeter spectrum that is rich in H$_{2}$O. Eight H$_{2}$O lines in emission and one in absorption were detected along with two H$_{2}^{18}$O lines in emission. The H$_{2}$O lines had upper state energies up to $E_{\mathrm{lower}}~600$~K.
   \end{itemize}

We have used multicomponent radiative transfer modeling to analyze the absorption-dominated PACS spectra together with the emission-dominated SPIRE spectrum and the continuum levels. The important conclusions from this work are: 
  \begin{itemize}
      \item Very high H$_{2}$ column density ($(0.3-1.0) \times 10^{25}$~cm$^{-2}$) towards the core of Zw~049.057 indicates that it is Compton-thick. This thick nuclear medium is also responsible for the frequency degradation of the intrinsic luminosity as high energy photons are absorbed and re-emitted at longer wavelengths.
      \item High H$_{2}$O column density per unit of continuum optical depth at $50$~$\mu$m of $\sim10^{18}$~cm$^{-2}$, an OH/H$_{2}$O ratio of $0.4-0.8$, and a dust temperature of $100-130$~K are derived towards the same nuclear region. The molecular abundance relative to H$_{2}$ in this region is also estimated to be high, $\chi_{\mathrm{H_2O}} \sim 5 \times 10^{-6}$.
      \item The high surface brightness of $(0.4-1.5) \times 10^{14}$~L$_{\sun}$~kpc$^{-2}$ on a spatial scale of $10-30$~pc is indicative of either a buried AGN or a very dense nuclear starburst.
      \item The absorption in the $^{18}$O isotopologue of OH indicates enhancement of $^{18}$O with a low $^{16}$O/$^{18}$O ratio of $50-100$. If this ratio is a tracer of starburst evolution it would place the galaxy in the same phase as Arp~220, in between the less evolved NGC~4418 and the more evolved Mrk~231.
   \end{itemize}

\begin{acknowledgements}
We thank the anonymous referee for a thorough and constructive report that helped improve the paper. NF and SA thank the Swedish National Space Board for generous grant support (grant numbers 145/11:1B, 285/12 and 145/11:1-3). Basic research in IR astronomy at NRL is funded by the US-ONR; J.F. acknowledges support from NHSC/ JPL subcontract 1456609.
\end{acknowledgements}

\bibliographystyle{aa}
\bibliography{ref}

\end{document}